\pdfoutput=1
\documentclass[12pt,a4paper]{article}

\usepackage{ifthen} 
\newboolean{pdflatex}
\setboolean{pdflatex}{true} 

\newboolean{articletitles}
\setboolean{articletitles}{true} 

\newboolean{uprightparticles}
\setboolean{uprightparticles}{false} 

\def\paperauthors{LHCb collaboration} 
\def\paperasciititle{Direct CP violation in charmless three-body decays of B mesons} 
\def\papertitle{Direct \CP violation in charmless three-body decays of $B^{\pm}$ mesons} 
\def\paperkeywords{{High Energy Physics}, {LHCb}} 
\def\papercopyright{\the\year\ CERN for the benefit of the LHCb collaboration} 
\def\paperlicence{CC BY 4.0 licence}
\def\paperlicenceurl{https://creativecommons.org/licenses/by/4.0/}

\usepackage[top=1in, bottom=1.25in, left=1in, right=1in]{geometry}

\usepackage{microtype}
\usepackage{lineno}  
\usepackage{xspace} 
\usepackage{caption} 

\usepackage{graphicx}  
\usepackage{color}
\usepackage{colortbl}
\graphicspath{{./figs/}} 

\usepackage{amsmath} 
\usepackage{amssymb}
\usepackage{amsfonts}
\usepackage{upgreek} 

\newcommand*\patchAmsMathEnvironmentForLineno[1]{%
\expandafter\let\csname old#1\expandafter\endcsname\csname #1\endcsname
\expandafter\let\csname oldend#1\expandafter\endcsname\csname
end#1\endcsname
 \renewenvironment{#1}%
   {\linenomath\csname old#1\endcsname}%
   {\csname oldend#1\endcsname\endlinenomath}%
}
\newcommand*\patchBothAmsMathEnvironmentsForLineno[1]{%
  \patchAmsMathEnvironmentForLineno{#1}%
  \patchAmsMathEnvironmentForLineno{#1*}%
}
\AtBeginDocument{%
\patchBothAmsMathEnvironmentsForLineno{equation}%
\patchBothAmsMathEnvironmentsForLineno{align}%
\patchBothAmsMathEnvironmentsForLineno{flalign}%
\patchBothAmsMathEnvironmentsForLineno{alignat}%
\patchBothAmsMathEnvironmentsForLineno{gather}%
\patchBothAmsMathEnvironmentsForLineno{multline}%
\patchBothAmsMathEnvironmentsForLineno{eqnarray}%
}

\usepackage{hyperxmp}

\usepackage[pdftex,
            pdfauthor={\paperauthors},
            pdftitle={\paperasciititle},
            pdfkeywords={\paperkeywords},
            pdfcopyright={Copyright (C) \papercopyright},
            pdflicenseurl={\paperlicenceurl}]{hyperref}

\usepackage[colorinlistoftodos,textsize=scriptsize]{todonotes}

\usepackage[bottom,flushmargin,hang,multiple]{footmisc}

\usepackage[all]{hypcap} 

\usepackage{xspace} 
\usepackage{upgreek}

\def\MagUp {\mbox{\em Mag\kern -0.05em Up}\xspace}

\ifthenelse{\boolean{uprightparticles}}%
{

 \def\Ppi         {\ensuremath{\uppi}\xspace}

 \def\Ppsi        {\ensuremath{\uppsi}\xspace}

 \def\PDelta      {\ensuremath{\Delta}\xspace}                 
 \def\PXi         {\ensuremath{\Xi}\xspace}                 
 \def\PLambda     {\ensuremath{\Lambda}\xspace}                 
 \def\PSigma      {\ensuremath{\Sigma}\xspace}                 
 \def\POmega      {\ensuremath{\Omega}\xspace}                 
 \def\PUpsilon    {\ensuremath{\Upsilon}\xspace}

 \def\PB      {\ensuremath{\mathrm{B}}\xspace}                 
                  
 \def\PD      {\ensuremath{\mathrm{D}}\xspace}

 \def\PJ      {\ensuremath{\mathrm{J}}\xspace}                 
 \def\PK      {\ensuremath{\mathrm{K}}\xspace}

 \def\Pd      {\ensuremath{\mathrm{d}}\xspace}

 \def\Ph      {\ensuremath{\mathrm{h}}\xspace}                 
 \def\Pi      {\ensuremath{\mathrm{i}}\xspace}

 \def\Ps      {\ensuremath{\mathrm{s}}\xspace}                 
                  
 \def\Pu      {\ensuremath{\mathrm{u}}\xspace}

 \def\thebaroffset{0.0em}
}
{

 \def\Ppi         {\ensuremath{\pi}\xspace}

 \def\Ppsi        {\ensuremath{\psi}\xspace}                 
                  
 \mathchardef\PDelta="7101
 \mathchardef\PXi="7104
 \mathchardef\PLambda="7103
 \mathchardef\PSigma="7106
 \mathchardef\POmega="710A
 \mathchardef\PUpsilon="7107
                  
 \def\PB      {\ensuremath{B}\xspace}                 
                  
 \def\PD      {\ensuremath{D}\xspace}

 \def\PJ      {\ensuremath{J}\xspace}                 
 \def\PK      {\ensuremath{K}\xspace}

 \def\Pd      {\ensuremath{d}\xspace}

 \def\Ph      {\ensuremath{h}\xspace}                 
 \def\Pi      {\ensuremath{i}\xspace}

 \def\Ps      {\ensuremath{s}\xspace}                 
                  
 \def\Pu      {\ensuremath{u}\xspace}

 \def\thebaroffset{0.18em}
}
\newcommand{\offsetoverline}[2][\thebaroffset]{\kern #1\overline{\kern -#1 #2}}%

\makeatletter
\ifcase \@ptsize \relax
  \newcommand{\miniscule}{\@setfontsize\miniscule{4}{5}}
\or
  \newcommand{\miniscule}{\@setfontsize\miniscule{5}{6}}
\or
  \newcommand{\miniscule}{\@setfontsize\miniscule{5}{6}}
\fi
\makeatother

\DeclareRobustCommand{\optbar}[1]{\shortstack{{\miniscule (\rule[.5ex]{1.25em}{.18mm})}
  \\ [-.7ex] $#1$}}

\def\uquark    {{\ensuremath{\Pu}}\xspace}

\def\dquark    {{\ensuremath{\Pd}}\xspace}

\def\squark    {{\ensuremath{\Ps}}\xspace}

\def\pion   {{\ensuremath{\Ppi}}\xspace}

\def\pip    {{\ensuremath{\pion^+}}\xspace}
\def\pim    {{\ensuremath{\pion^-}}\xspace}
\def\pipm   {{\ensuremath{\pion^\pm}}\xspace}

\def\kaon    {{\ensuremath{\PK}}\xspace}

\def\KorKbar {\kern \thebaroffset\optbar{\kern -\thebaroffset \PK}{}\xspace}

\def\Kp      {{\ensuremath{\kaon^+}}\xspace}
\def\Km      {{\ensuremath{\kaon^-}}\xspace}
\def\Kpm     {{\ensuremath{\kaon^\pm}}\xspace}

\def\Dbar    {{\ensuremath{\offsetoverline{\PD}}}\xspace}
\def\D       {{\ensuremath{\PD}}\xspace}

\def\DorDbar {\kern \thebaroffset\optbar{\kern -\thebaroffset \PD}\xspace}
\def\Dz      {{\ensuremath{\D^0}}\xspace}
\def\Dzb     {{\ensuremath{\Dbar{}^0}}\xspace}
\def\Dp      {{\ensuremath{\D^+}}\xspace}
\def\Dm      {{\ensuremath{\D^-}}\xspace}

\def\DpDm    {\ensuremath{\Dp {\kern -0.16em \Dm}}\xspace}

\def\B       {{\ensuremath{\PB}}\xspace}

\def\BorBbar {\kern \thebaroffset\optbar{\kern -\thebaroffset \PB}\xspace}

\def\Bd      {{\ensuremath{\B^0}}\xspace}

\def\BdorBdbar {\kern \thebaroffset\optbar{\kern -\thebaroffset \Bd}\xspace}
\def\Bu      {{\ensuremath{\B^+}}\xspace}
\def\Bub     {{\ensuremath{\B^-}}\xspace}
\def\Bp      {{\ensuremath{\Bu}}\xspace}
\def\Bm      {{\ensuremath{\Bub}}\xspace}
\def\Bpm     {{\ensuremath{\B^\pm}}\xspace}

\def\Bs      {{\ensuremath{\B^0_\squark}}\xspace}

\def\BsorBsbar {\kern \thebaroffset\optbar{\kern -\thebaroffset \Bs}\xspace}

\def\jpsi     {{\ensuremath{{\PJ\mskip -3mu/\mskip -2mu\Ppsi}}}\xspace}

\def\Y#1S{\ensuremath{\PUpsilon{(#1S)}}\xspace}

\def\LorLbar     {\kern \thebaroffset\optbar{\kern -\thebaroffset \PLambda}\xspace}

\newcommand{\decay}[2]{\ensuremath{#1\!\to #2}\xspace} 

\def\to                 {\ensuremath{\rightarrow}\xspace}

\def\CP                {{\ensuremath{C\!P}}\xspace}

\def\Bhhh        {\decay{\Bpm}{\Ph^{\pm} \Ph'^+\Ph'^-}}
\def\Bjpsik      {\decay{\Bpm}{\jpsi\Kpm}}

\def\Bkpp        {\decay{\Bpm}{\Kpm \pip\pim}}

\def\AT#1     {\ensuremath{A_{\mathrm{T}}^{#1}}\xspace}           

\def\C#1      {\ensuremath{\mathcal{C}_{#1}}\xspace}                       
\def\Cp#1     {\ensuremath{\mathcal{C}_{#1}^{'}}\xspace}                    
\def\Ceff#1   {\ensuremath{\mathcal{C}_{#1}^{\mathrm{(eff)}}}\xspace}        
\def\Cpeff#1  {\ensuremath{\mathcal{C}_{#1}^{'\mathrm{(eff)}}}\xspace}       
\def\Ope#1    {\ensuremath{\mathcal{O}_{#1}}\xspace}                       
\def\Opep#1   {\ensuremath{\mathcal{O}_{#1}^{'}}\xspace}                    

\newcommand{\aunit}[1]{\ensuremath{\text{\,#1}}}       

\newcommand{\tev}{\aunit{Te\kern -0.1em V}\xspace}
\newcommand{\gev}{\aunit{Ge\kern -0.1em V}\xspace}
\newcommand{\mev}{\aunit{Me\kern -0.1em V}\xspace}
\newcommand{\kev}{\aunit{ke\kern -0.1em V}\xspace}
\newcommand{\ev}{\aunit{e\kern -0.1em V}\xspace}
 
\newcommand{\mevc}{\ensuremath{\aunit{Me\kern -0.1em V\!/}c}\xspace}
\newcommand{\gevc}{\ensuremath{\aunit{Ge\kern -0.1em V\!/}c}\xspace}
\newcommand{\mevcc}{\ensuremath{\aunit{Me\kern -0.1em V\!/}c^2}\xspace}
\newcommand{\gevcc}{\ensuremath{\aunit{Ge\kern -0.1em V\!/}c^2}\xspace}
\newcommand{\gevgevcccc}{\ensuremath{\gev^2\!/c^4}\xspace} 

\def\fb   {\ensuremath{\aunit{fb}}\xspace}
\def\invfb   {\ensuremath{\fb^{-1}}\xspace}

\def\gsim{{~\raise.15em\hbox{$>$}\kern-.85em
          \lower.35em\hbox{$\sim$}~}\xspace}
\def\lsim{{~\raise.15em\hbox{$<$}\kern-.85em
          \lower.35em\hbox{$\sim$}~}\xspace}

\def\sPlot{\mbox{\em sPlot}\xspace}

\def\pt         {\ensuremath{p_{\mathrm{T}}}\xspace}

\def\evtgen     {\mbox{\textsc{EvtGen}}\xspace}

\def\geant      {\mbox{\textsc{Geant4}}\xspace}

\def\photos     {\mbox{\textsc{Photos}}\xspace}

\def\pythia     {\mbox{\textsc{Pythia}}\xspace}

\def\tell1  {TELL1\xspace}
\def\ukl1   {UKL1\xspace}

\newcommand{\lhcborcid}[1]{\href{https://orcid.org/#1}{\hspace*{0.1em}\raisebox{-0.45ex}{\includegraphics[width=1em]{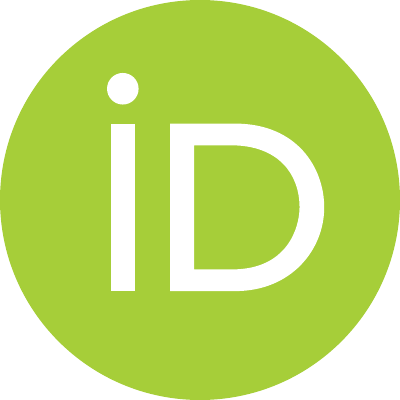}}}}

\def\pipipi {\ensuremath{{\Bpm \to \pipm \pip \pim}}\xspace}
\def\kpipi {\ensuremath{{\Bpm \to \Kpm \pip \pim}}\xspace}

\def\kkpi {\ensuremath{{\Bpm \to \pipm \Kp \Km }}\xspace}
\def\kkk {\ensuremath{{\Bpm \to \Kpm \Kp \Km}}\xspace}

\def\jpsik {\ensuremath{{\Bpm \to \jpsi \Kpm}}\xspace}

\def\acp {\ensuremath{A_{\CP}}\xspace}

\def\acpraw {\ensuremath{A_{\rm raw}}\xspace}

\def\acprawacc {\ensuremath{A_{\rm raw}^{\rm corr}}\xspace}

\def\aprod {\ensuremath{A_{\rm P}}\xspace}

\newcommand{\approptoinn}[2]{\mathrel{\vcenter{
  \offinterlineskip\halign{\hfil$##$\cr
    #1\propto\cr\noalign{\kern2pt}#1\sim\cr\noalign{\kern-2pt}}}}}

\usepackage{cite} 
\usepackage{mciteplus}

\usepackage{longtable} 

\usepackage{overpic}
\usepackage{amsmath}
\usepackage{placeins}

\usepackage{float}
\usepackage{adjustbox}
\usepackage{graphicx}

\usepackage{microtype}
\usepackage{lineno}  
\usepackage{xspace} 
\usepackage{caption} 

\usepackage{cite} 
\usepackage{mciteplus}

\begin{document}

\renewcommand{\thefootnote}{\fnsymbol{footnote}}
\setcounter{footnote}{1}


\begin{titlepage}
\pagenumbering{roman}

\vspace*{-1.5cm}
\centerline{\large EUROPEAN ORGANIZATION FOR NUCLEAR RESEARCH (CERN)}
\vspace*{1.5cm}
\noindent
\begin{tabular*}{\linewidth}{lc@{\extracolsep{\fill}}r@{\extracolsep{0pt}}}
\ifthenelse{\boolean{pdflatex}}
{\vspace*{-1.5cm}\mbox{\!\!\!\includegraphics[width=.14\textwidth]{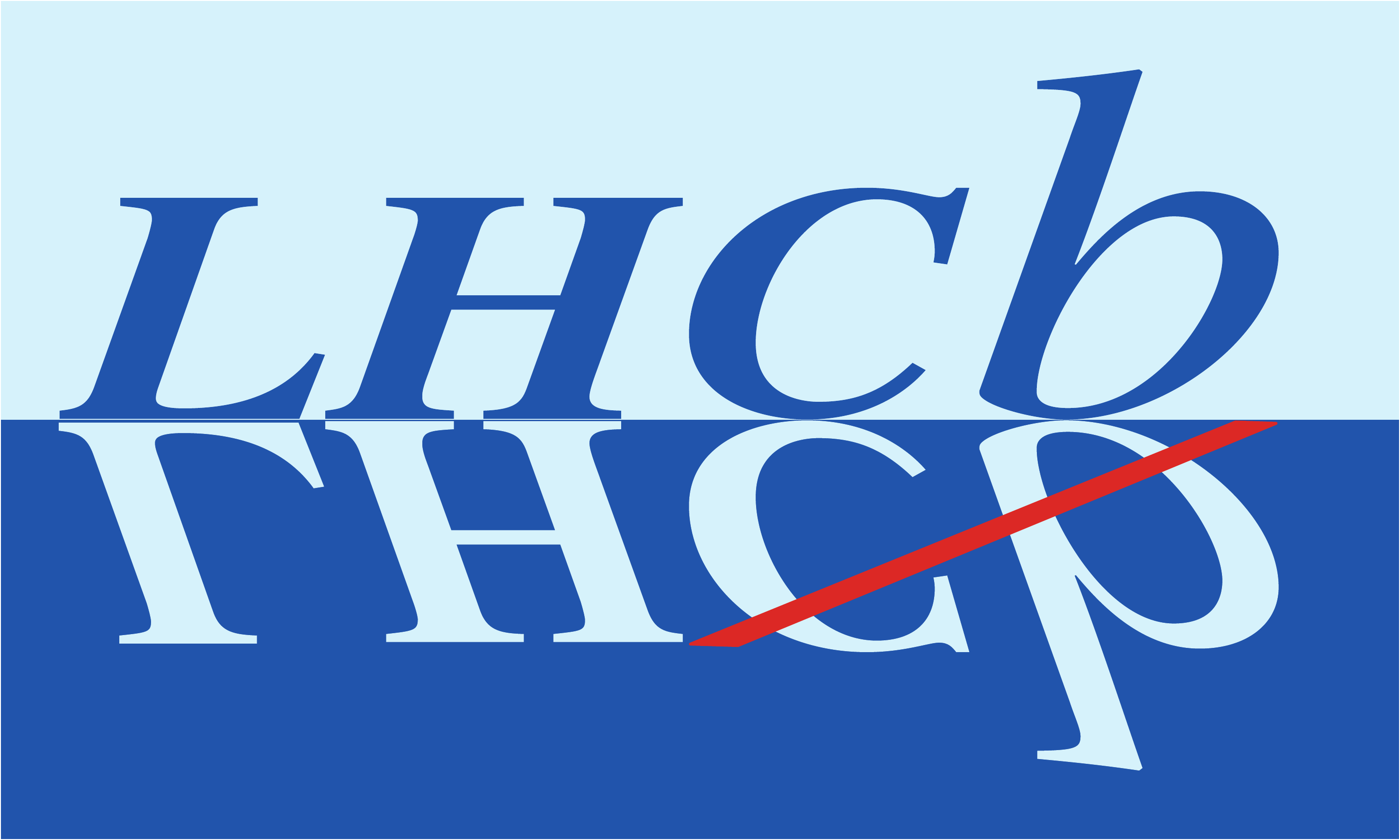}} & &}%
{\vspace*{-1.2cm}\mbox{\!\!\!\includegraphics[width=.12\textwidth]{figs/lhcb-logo.eps}} & &}%
\\
 & & CERN-EP-2022-100 \\  
 & & LHCb-PAPER-2021-049 \\  
 & & \today \\ 
 & & \\
\end{tabular*}

\vspace*{3.0cm}

{\normalfont\bfseries\boldmath\huge
\begin{center}
  \papertitle 
\end{center}
}

\vspace*{2.0cm}

\begin{center}
\paperauthors\footnote{Authors are listed at the end of this paper.}
\end{center}

\vspace{\fill}

\begin{abstract}
\noindent Measurements of \CP asymmetries in charmless three-body decays of $B^\pm$ mesons are reported using proton-proton collision data collected by the LHCb detector, corresponding to an integrated luminosity of 5.9\invfb. 
The previously observed \CP asymmetry in \kkpi decays is confirmed, and \CP asymmetries are observed with a significance of more than five standard deviations in the \pipipi and \kkk decays, while the \CP asymmetry of \kpipi decays is confirmed to be compatible with zero. The distributions of these asymmetries are also studied as a function of the three-body phase space and suggest contributions from rescattering and resonance interference processes. An indication of the presence of the decays $B^\pm \to \pi^\pm \chi_{c0}(1P)$ in both \pipipi and \kkpi decays is observed, as is \CP violation involving these amplitudes.
  
\end{abstract}

\vspace*{2.0cm}

\begin{center}
  Published in Phys. Rev. D108 (2023) 012008
\end{center}

\vspace{\fill}

{\footnotesize 
\centerline{\copyright~\papercopyright. \href{\paperlicenceurl}{\paperlicence}.}}
\vspace*{2mm}

\end{titlepage}


\newpage
\setcounter{page}{2}
\mbox{~}
%
%
%
%


\renewcommand{\thefootnote}{\arabic{footnote}}
\setcounter{footnote}{0}

\cleardoublepage


\pagestyle{plain} 
\setcounter{page}{1}
\pagenumbering{arabic}



\section{Introduction}
\label{sec:introduction}

The violation of charge-parity (\CP) symmetry has been observed in $B^+$, $B^0$ and $B^0_s$ decays, and yet our understanding of the dynamics involved in these asymmetries is incomplete. Some crucial questions remain to be understood: what is the dynamical origin of the weak phase associated to the decay amplitudes? Why has \CP violation not been observed so far in baryon decays? What is the primary source of the strong phase difference in the observed \CP asymmetries~\cite{BediagaGobel}?

The last question is directly related to the longstanding debate about the role of short- and long-distance contributions to the
generation of the strong-phase differences, needed for direct \CP violation to occur, and three-body decays of $B$ mesons offer a way of finding the answer. 
A recent amplitude analysis found a large \CP asymmetry related to the interference between the \textit{S}- and \textit{P}-wave contributions in \pipipi decays~\cite{LHCb-PAPER-2019-017,LHCb-PAPER-2019-018}. 
In a similar  analysis, large \CP violation involving $\pi\pi \to KK$ rescattering was observed in \kkpi decays~\cite{LHCb-PAPER-2018-051}.
The patterns of localized \CP asymmetries in the four charmless decay modes \pipipi, \kkk, \kkpi and \kpipi, reported in Refs.~\cite{LHCb-PAPER-2013-051,LHCb-PAPER-2014-044}, can be interpreted as originating from long-distance hadronic interactions~\cite{BediagaGobel}.

Heavy meson three-body decays can be understood in terms of a superposition of intermediate states involving resonances. From this perspective, the phase-space-integrated \CP asymmetries (\acp) observed in charmless three-body decays of $B^{\pm}$ decays are the sum of the asymmetries of each intermediate state. In the \kkpi decay~\cite{LHCb-PAPER-2014-044}, the observed \CP asymmetry would result predominantly from the $\pi\pi \to KK$ rescattering amplitude, weighted by its fit fraction~\cite{LHCb-PAPER-2018-051}. 
The same can be seen from the recent amplitude analysis of the \pipipi decay \cite{LHCb-PAPER-2019-017,LHCb-PAPER-2019-018}. The sum of the \CP asymmetries for the $B^{\pm} \to \sigma \pi^{\pm}$ and  $B^{\pm} \to f_2(1270) \pi^{\pm}$ decays weighted by their respective fractions is almost equal to the inclusive \CP asymmetry observed for the \pipipi decay~\cite{LHCb-PAPER-2014-044}.

Another approach considers the global properties of the three-body amplitudes.
The authors of Refs.~\cite{Gronau2003,Gronau2013,Gronau2014} use U-spin symmetry to relate the \CP asymmetries and partial decay widths of $B^\pm \to h^\pm h'^+ h'^-$ decays. U-spin symmetry predicts that the ratios \mbox{$\Delta\Gamma(B^\pm \to K^\pm \pi^+ \pi^-) / \Delta\Gamma(B^\pm \to \pi^\pm K^+ K^-)$} and \mbox{$\Delta\Gamma(B^\pm \to K^\pm K^+ K^-) / \Delta\Gamma(B^\pm \to \pi^\pm \pi^+ \pi^-)$}, where $\Delta \Gamma$ is the partial decay width difference between $B^-$ and $B^+$, must be close to $-1$. These predictions, based on a general three-body decay approach, are supported by experimental measurements, within the  uncertainties~\cite{LHCb-PAPER-2014-044, Gronau2013,Xu2013, Pat2021}.

These two approaches are complementary and the role of each one must be determined from experimental results, such as the \CP asymmetries and the branching fractions of charmless three-body \Bpm decays, recently presented by the LHCb collaboration~\cite{LHCb-PAPER-2020-031}.

In this article, \CP violation is studied in four charmless $B^{\pm}$ meson decays into three charged pseudoscalar particles: \kpipi, \kkk, \kkpi and \pipipi. 
The analysis is based on proton-proton ($pp$) collision data at a center-of-mass energy of 13\tev, corresponding to an integrated luminosity of 5.9\invfb, recorded with the LHCb detector between 2015 and 2018. 
Phase-space integrated \CP asymmetries are measured, as well as \CP asymmetries in specific regions of the Dalitz plots. The analysis uses the decay $B^\pm \to J/\psi(\to \mu^+ \mu^-) K^\pm$ as a control channel, in order to take into account the production asymmetry of the $B^{\pm}$ mesons.


\section{LHCb detector, dataset and simulation}
\label{sec:detector}

The LHCb detector~\cite{LHCb-DP-2008-001, LHCb-DP-2014-002} 
is a single-arm forward spectrometer covering the pseudorapidity range $2 < \eta < 5$, designed for the study of particles containing $b$ or $c$ quarks. The detector includes a high-precision tracking system consisting of a silicon-strip vertex  detector surrounding the $pp$ interaction region, a large-area silicon-strip detector located upstream of a dipole magnet with a bending power of about 4 Tm, and three stations of silicon-strip detectors and straw drift tubes placed downstream of the magnet. The tracking system provides a measurement of the momentum, $p$, of charged particles with a relative uncertainty that varies from 0.5\% at low momentum ($\sim$ 2\gevc) to 1.0\% at 200\gevc. 
The minimum distance of a track to a primary $pp$ collision vertex (PV), the impact parameter (IP), is measured with a resolution of (15 + 29/\pt)\,$\upmu$m, where \pt is the component of the momentum transverse to the beam, in\,\gevc. Different types of charged hadrons are distinguished using information from two ring-imaging Cherenkov detectors. The corresponding efficiencies and misidentification rates are given in Ref.\cite{Calabrese:2811056}. Photons, electrons and hadrons are identified by a calorimeter system consisting of scintillating-pad and preshower detectors, an electromagnetic and a hadronic calorimeter. Muons are identified by a system composed of alternating layers of iron and multiwire proportional chambers.

The online event selection is performed by a trigger, which consists of a hardware stage, based on information from the calorimeter system, followed by a software stage, which applies a full event reconstruction. 
At the hardware trigger stage, events are selected if a transverse energy, typically larger than 3.5\gev, is deposited in the calorimeters by at least one hadron from the $B^{\pm}\to h^{\pm}h'^+h'^-$ candidates.
For the control channel, it is required that the hardware trigger decision must also be due to at least one muon with high \pt. In all final states, events are also selected if the trigger decision is independent of the signal.

The software trigger requires a two-, three- or four-track vertex with a significant displacement from any PV. At least one charged particle must have a large \pt and be inconsistent with originating from any PV\@. A multivariate algorithm is used for the identification of displaced vertices consistent with the decay of a {\it b}-hadron \cite{LHCb-DP-2019-001}.

Samples of simulated events are used to model the effects of the detector efficiency and the selection requirements, to determine the fit models and to evaluate efficiencies. In the simulation, $pp$ collisions are generated using \pythia 8 \cite{Sj_strand_2015} with a specific LHCb configuration \cite{LHCb-PROC-2010-056}. Decays of unstable particles are described by \evtgen \cite{Lange:2001uf}, in which final-state radiation is generated using \photos \cite{davidson2015photos}. The interaction of the generated particles with the detector, and its response, are implemented using the \geant toolkit \cite{Agostinelli:2002hh,Allison:2006ve} as described in Ref. \cite{LHCb-PROC-2011-006}.


\section{Candidate selection}
\label{sec:selection}

The selection of the four \Bhhh samples is based on simulations and data-driven methods. The analyzed  decay modes share the same topology and similar kinematics, leading to a common strategy for the selection. Requirements are made on the quality of the three-track vertices: the decay vertex must be detached from the PV, and the momentum vector of the reconstructed \Bpm candidate must point back to the PV\@. The final-state particles must have a significant IP with respect to the PV\@.

The combinatorial background is reduced by a multivariate analysis of a set of ten topological and kinematic variables, using boosted decision trees (BDT)~\cite{Breiman} as classifiers. 
Samples of simulated events are used to describe the signal, whereas candidates satisfying $m(h^\pm h^{'+} h^{'-})>5.4\gevcc$, are used to represent the combinatorial background. The BDT classifier is trained separately for each decay mode. 
A requirement on the BDT response for each mode is chosen to maximise the ratio $\varepsilon_{\textrm{sim}}/\sqrt{(S+B)_{\textrm{data}}}$, where $\varepsilon_{\textrm{sim}}$ is the signal efficiency from simulation and $(S+B)_{\textrm{data}}$ is obtained by counting the candidates from data with invariant mass in the interval of $\pm$40\mevcc centred on the known $B^+$ mass~\cite{PDG2020}. The most discriminating BDT variables are related to the $B^{\pm}$ candidate distance of closest approach to the PV, pointing angle, decay vertex quality, and IP of each final-state particle.

Particle identification (PID) requirements are made on all three final-state particles, reducing the contamination from other \Bhhh decays to the per cent level. This background is due mainly to $K-\pi$  misidentification. 
The analysis is restricted to particles with momentum below 100\gevc, since above this limit the discrimination power of the PID system is limited. In order to avoid the contamination from semileptonic decays, a muon veto is placed on all decay products by applying requirements on track reconstruction and trigger decisions. Finally, a method based on Artificial Neural Network is employed to remove electrons~\cite{LHCb-DP-2018-001}. 

More stringent PID requirements are applied to the \mbox{\pipipi} and \mbox{\kkpi} decays. These decays have lower branching fractions compared to the \mbox{\kpipi} and \mbox{\kkk} decays and are therefore more severely affected by the cross-feed from other \Bhhh decays.

There are significant backgrounds from two-body decays of \Bpm mesons into a \Dz or \Dzb meson and a pseudoscalar particle, followed by the two-body decay of the \Dz meson into pions and kaons. This background is removed by excluding candidates for which the two-body invariant masses $m(\pi^+\pi^-)$, $m(K^{\pm}\pi^{\mp})$ or $m(K^+K^-)$ are within  $\pm30$\mevcc of the known \Dz mass~\cite{PDG2020}.

After the final selection, approximately 0.5\% of the events contain more than one signal candidate. One candidate is chosen randomly in these cases.

The decay \Bjpsik, with subsequent decay \jpsi$\to \mu^+\mu^-$ is used to measure the \Bpm production asymmetry. The sample is selected with the same requirements as the \Bkpp decay, with three exceptions: the particle identification, in which the pion PID is substituted by that of muon; the use of events for which the hardware trigger decision is due to a muon with high \pt; the requirement of the \jpsi candidate mass within the interval $3.05<m(\mu^+\mu^-)<3.15\gevcc$.


\section{Invariant mass fits}
\label{sec:massfits}

\begin{figure}[tb]
\begin{overpic}[width=0.49\linewidth]{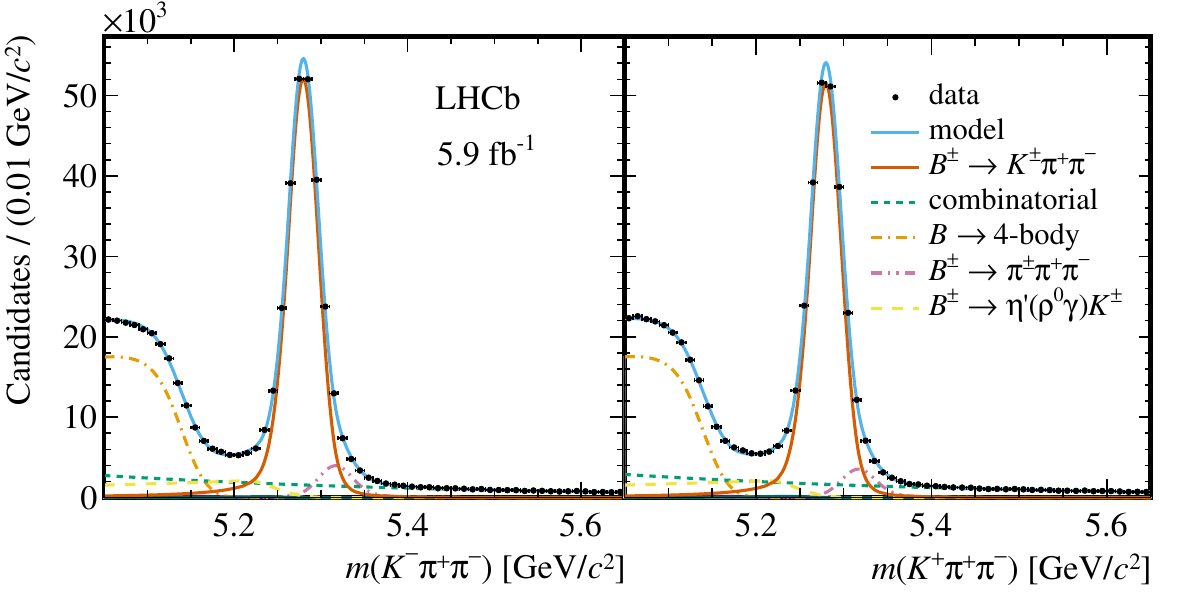}
 \put(11,44){\scriptsize{(a)}}
\end{overpic}
\begin{overpic}[width=0.47\linewidth]{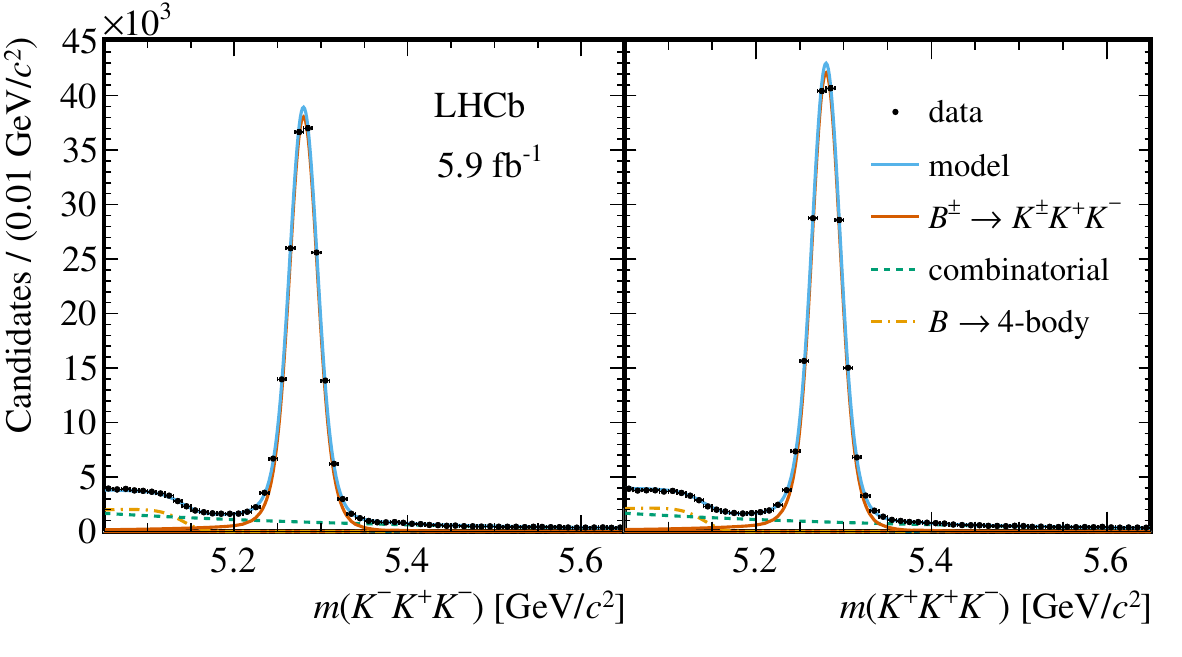}
 \put(11,47){\scriptsize{(b)}}
\end{overpic}\\
\begin{overpic}[width=0.49\linewidth]{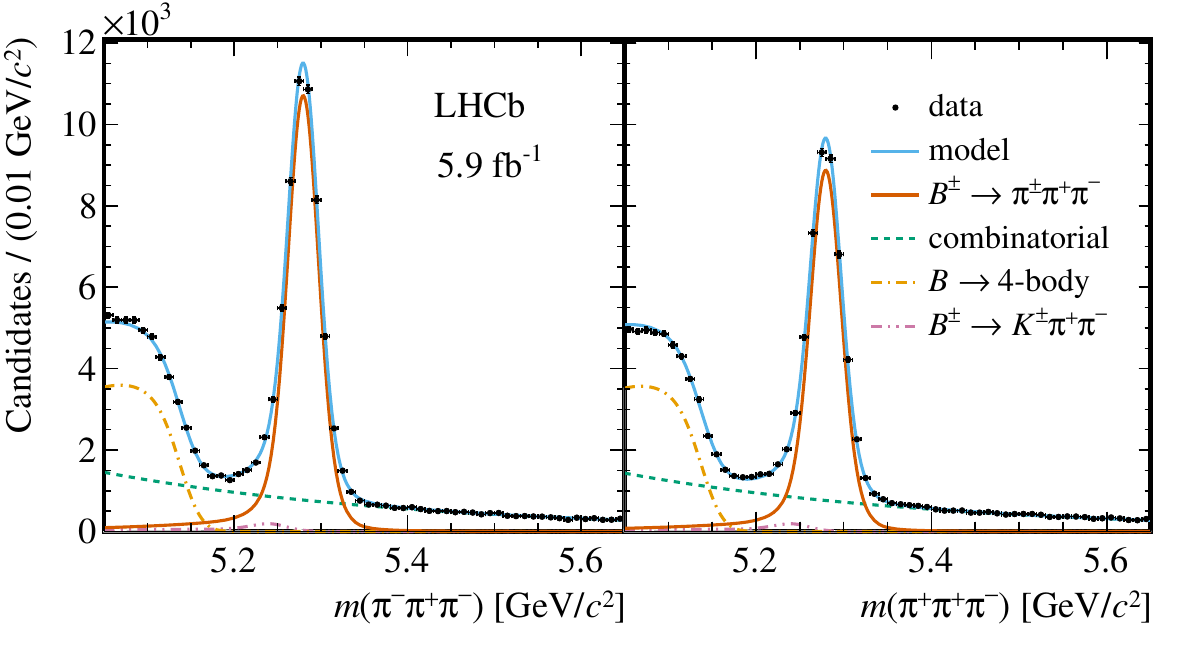}
 \put(11,47){\scriptsize{(c)}}
\end{overpic}
\begin{overpic}[width=0.49\linewidth]{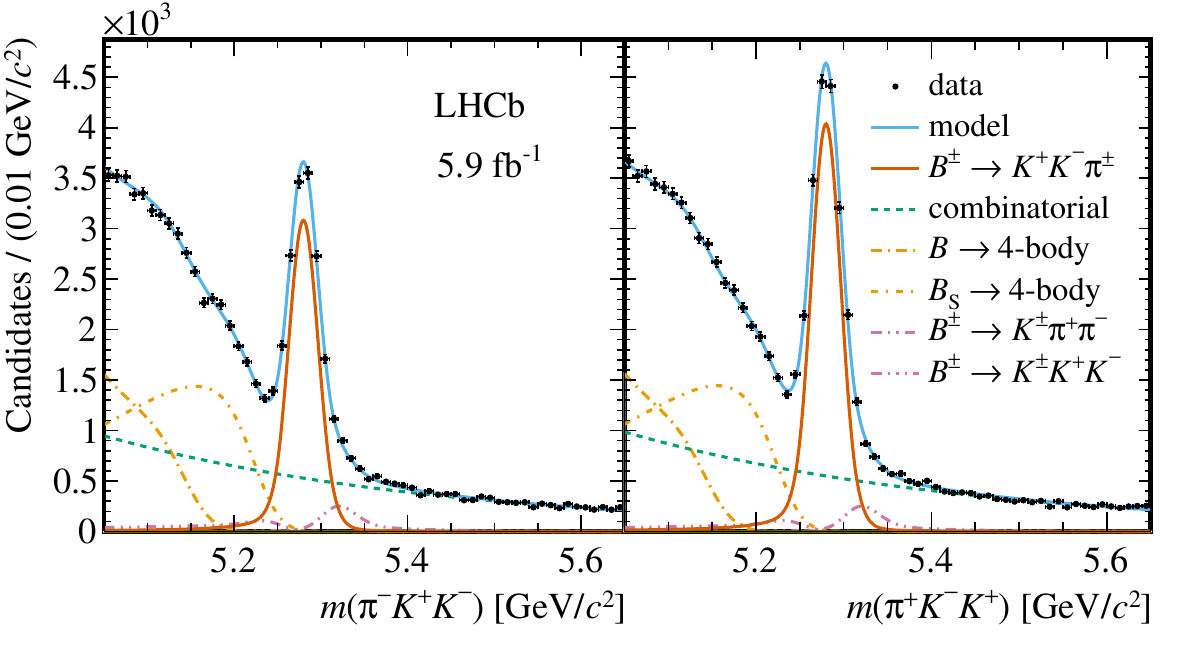}
  \put(11,47){\scriptsize{(d)}}
\end{overpic}
\caption{Invariant mass spectra of (a)\,\kpipi, (b)\,\kkk, (c)\,\pipipi and (d)\,\kkpi decays in the mass range [5050--5650]\mevcc.
The left panel in each figure shows the \Bm candidates and the right panel shows the \Bp candidates.
The results of the unbinned maximum likelihood fits are overlaid.
}
\label{fig:MassFit}
\end{figure}

The yield and raw asymmetry of each decay channel are extracted from simultaneous unbinned extended maximum-likelihood fits to the \Bm and \Bp invariant mass distributions, as shown in Fig.~\ref{fig:MassFit}. 
The signal components in each of the four channels are parameterized by a sum of a Gaussian and two Crystal Ball functions~\cite{das2016simple} with a common mean and different widths and tails on either side of the peak, accounting for asymmetric effects such as quark final-state radiation. 
The combinatorial backgrounds is described by an exponential function. The slopes and charge asymmetry of the combinatorial background are free parameters in the fit, the latter being compatible with zero when systematic uncertainties, production and detection asymmetries are taken into account.
The background due to partially reconstructed four-body \B decays, in which one particle is not reconstructed, is parameterized by an ARGUS function~\cite{Argus} convolved with a Gaussian resolution function. A contribution from four-body \Bs decays is also present in the \kkpi sample. 

The dominant peaking backgrounds in the signal regions arise from other three-body \B decays with one or more misidentified final-state particles. 
For \kpipi decays, the largest contribution comes from \pipipi with the pion misidentified as kaon, followed by the partially reconstructed decay $\Bpm\to \eta'(\rho^{0} \gamma)\Kpm $, in which the photon is undetected. 
Conversely, the only peaking contribution to the \pipipi spectrum comes from \kpipi decays. 
The peaking background to \kkk is the result of single or double misidentification of pions as kaons from \kkpi and \kpipi decays. 
The \kkpi decay has the largest number of significant peaking background contributions due to its lower branching fraction and both $K\to\pi$ and $\pi\to K$ misidentification possibilities, which is why the three other channels can contribute to its spectrum. 

The shapes and fractions of peaking backgrounds relative to the signal yield are obtained from simulation of the relevant decay modes and fixed in the fits, whereas the yields of the partially reconstructed background components vary freely in the fit. The resulting signal yields and raw asymmetries ($A_{\text{raw}}$) are listed in Table~\ref{tab:signalyields}.
The data samples are larger than those presented in Ref.~\cite{LHCb-PAPER-2014-044} due to both an increase in the integrated luminosity and the production cross-section, and the use of a more efficient selection.

\begin{table}[tb]
\caption{Total signal yields, raw asymmetries and average efficiency ratio $R$ of charmless three-body \Bpm decays for the inclusive data set.}
\centering
\begin{tabular}{lccc}
\hline
Decay mode & Total yield &  $A_{\text{raw}}$ & $R = \,<\epsilon^->/<\epsilon^+>$ \\ \hline

\kpipi     & $499\,200 \pm 900 \phantom{0}$            & $+0.006 \pm 0.002$ & $1.0038 \pm 0.0027$ \\
\kkk       & $365\,000 \pm 1000 $                      & $-0.052 \pm 0.002$ & $0.9846 \pm 0.0024$\\
\pipipi    & $101\,000 \pm 500 \phantom{0}$            & $+0.090 \pm 0.004$ & $1.0354 \pm 0.0037$\\
\kkpi      & $\phantom{3} 32\,470 \pm 300 \phantom{0}$ & $-0.132 \pm 0.007$ & $0.9777 \pm 0.0032$ \\ \hline
\end{tabular}
\label{tab:signalyields}
\end{table}


\section{\boldmath Phase-space integrated \CP asymmetries}
\label{sec:global}

Neither the selection efficiencies nor the observed raw asymmetries are uniform across the Dalitz plot, and efficiencies may differ for \Bp and \Bm.
Therefore, the efficiency of selecting a signal decay is parameterized in the two-dimensional square Dalitz plot separately for \Bp and \Bm.
Such a transformation improves the resolution in the regions with high sensitivity by expanding the corners and the borders of the nominal Dalitz plot relative to the sparsely populated central region.
These maps are primarily obtained using simulated decays; however, effects arising from the hardware trigger and PID efficiency are determined using data calibration samples. The efficiency shows a variation from 0.9\% to 1.7\% for \kkk, 0.1\% to 0.6\% for \kkpi, 0.1\% to 1\% for \kpipi and 0.1\% to 1.2\% for \pipipi.
The efficiency correction applied to the integrated raw asymmetries is determined from the ratio between the \Bm and \Bp average efficiencies, as listed in Table~\ref{tab:signalyields}. 
These averages reflect the data distributions across the phase space by weighting the efficiency maps by the data events.

The phase-space integrated \CP asymmetries \acp are obtained by correcting the raw asymmetry for selection efficiency effects to get the efficiency-corrected raw asymmetries \acprawacc, and for the production  asymmetry $\aprod$. The efficiency-corrected raw asymmetry is defined as
\begin{equation}
\acprawacc \equiv \dfrac{N_{B^-}^{\rm corr} - N_{B^+}^{\rm corr}}{N_{B^-}^{\rm corr} + N_{B^+}^{\rm corr}} = \dfrac{(N_{B^-}/R) - N_{B^+}}{(N_{B^-}/R) + N_{B^+}}, 
\label{eq:acprawaccdef}
\end{equation}
where $N_{B^-}^{\rm corr}$ and $N_{B^+}^{\rm corr}$ are the number of \Bm and \Bp signal events obtained from the invariant mass fit, $N_{B^-}$ and $N_{B^+}$, corrected for the efficiency. 
They are related to the asymmetries by
\begin{eqnarray}
N_{\Bm}^{\rm corr} &=& \dfrac{N_{\Bm}^{\rm corr}+N_{\Bp}^{\rm corr}}{2} (1+\acp)(1+\aprod)\, , \label{eq:nbmacc} \\
N_{\Bp}^{\rm corr} &=& \dfrac{N_{\Bm}^{\rm corr}+N_{\Bp}^{\rm corr}}{2} (1-\acp)(1-\aprod).
\label{eq:nbpacc}  
\end{eqnarray}
Substituting Eqs.~\eqref{eq:nbmacc} and~\eqref{eq:nbpacc} into Eq.~\eqref{eq:acprawaccdef}, the physical \CP asymmetry can be expressed in terms of \acprawacc and \aprod:
\begin{equation}
\acp = \frac{ \acprawacc - \aprod }{ 1 - \acprawacc \aprod }.
\label{eq:acpNoAdet}
\end{equation} 

To obtain \aprod, the decay \jpsik is used as a control channel and its raw asymmetry is obtained from a fit to the invariant mass distribution. This decay has larger signal yield and lower background contributions compared to \Bhhh decays. 
Its background is parameterized as the sum of a combinatorial component, represented by a single exponential, and partially-reconstructed $B\to \jpsi K^*(892)^{\pm,0}$ decay, described by an ARGUS function. 
The \jpsik model is parameterized with the same function used for the \Bhhh channels. 
Systematic uncertainties of the raw asymmetry are obtained by varying the signal fit model, leaving the background asymmetry to vary in the fit, and looking at variations from different trigger samples of the data. The total systematic uncertainty is taken as the sum in quadrature of the individual uncertainties. 
The raw asymmetry of the control channel is measured to be $\acpraw(\jpsik) =-0.0118 \pm 0.0008 \,^{+0.0007}_{-0.0008}$, where the first uncertainty is statistical and the second systematic.

To obtain the $B^{\pm}$ production asymmetry, this raw asymmetry is corrected by its efficiency ratio, calculated using a sample of simulated events produced without any $B^{\pm}$ production asymmetry, to obtain \acprawacc as before, and the world average value of the \jpsik\ \CP asymmetry, $0.0018 \pm 0.0030$~\cite{PDG2020}, is subtracted
\begin{equation}
\aprod = \acprawacc(\jpsik)-\acp(\jpsik).
\label{eq:aprod}
\end{equation} 
The measured \B meson production asymmetry is \mbox{$\aprod = -0.0070 \pm 0.0008\,^{+0.0007}_{-0.0008} \pm 0.0030$}, where the last uncertainty is due to the \CP asymmetry of \jpsik decays~\cite{PDG2020}. 

Finally, the \CP asymmetries of the four \Bhhh modes are measured to be
\begin{align}
\acp(\kpipi)  = +0.011 \pm 0.002, \nonumber \\
\acp(\kkk)    = -0.037 \pm 0.002, \nonumber \\
\acp(\pipipi) = +0.080 \pm 0.004, \nonumber \\
\acp(\kkpi)   = -0.114 \pm 0.007, \nonumber
\end{align}
where the statistical uncertainties are  obtained from propagation of Eq.~\eqref{eq:acpNoAdet}, assuming no correlation term. 


\section{Systematic uncertainties and results}
\label{sec:systematics}

Several sources of systematic uncertainties are considered and can be broadly divided into three groups: potential mismodelling of the invariant mass distributions, phase-space efficiency corrections and knowledge of the $B^{\pm}$ production asymmetry. The systematic uncertainties due to the mass fit models are quantified by taking the difference in the \CP asymmetry resulting from variations of the model. The alternative fits have good quality and describe the data accurately. To estimate the uncertainty due to the choice of the signal mass function, the initial model is replaced by an alternative empirical distribution \cite{das2016simple}. 

The contribution associated with the peaking background fractions reflects the uncertainties in the expected yields determined from simulation and it is evaluated by varying the fractions within their statistical uncertainties. 
In addition, the systematic uncertainty associated to the fact that the peaking background asymmetry is fixed to zero is estimated by setting it to the value obtained in the previous analysis \cite{LHCb-PAPER-2014-044}, within the corresponding uncertainties. 
The uncertainty due to the choice of an exponential function to model the combinatorial component is estimated by repeating the fit using a second order polynomial function.

The systematic uncertainty related to the efficiency correction procedure consists of two parts: the statistical uncertainty in the detection efficiency due to the finite size of the simulated samples, and the uncertainty due to the finite size of the bins, which is evaluated by varying the binning used in the efficiency correction. The former term dominates for all decays.

The systematic uncertainty due to the $B^{\pm}$ production asymmetry measurement using the \jpsik control channel is obtained by summing in quadrature the experimental statistical and systematic uncertainties of \aprod.

\begin{table}[tb]
{\small
\begin{center}
\caption{Individual components and total systematic uncertainties related to the mass fit model, efficiency corrections for the integrated phase space and $B^{\pm}$ production asymmetry. The total is the sum in quadrature of these components.}
\begin{tabular}{lcccc}
\hline
Source of uncertainty & $K^{\pm}\pi^+\pi^-$ & $K^{\pm}K^+K^-$ & $\pi^{\pm}\pi^+\pi^-$ & $\pi^{\pm}K^+K^-$ \\ \hline
Signal model                 & 0.0004  & 0.0007 & 0.0000  & 0.0001 \\
Peaking background fraction  & 0.0005  & 0.0010 & 0.0002  & 0.0004 \\
Peaking background asymmetry & 0.0022  & 0.0001 & 0.0005  & 0.0007 \\
Combinatorial model          & 0.0002  & 0.0005 & 0.0015  & 0.0025 \\
Efficiency correction        & 0.0014  & 0.0016 & 0.0018  & 0.0019 \\ 
Production asymmetry           & 0.0011  & 0.0011 & 0.0011  & 0.0011 \\
\hline
Total                        & 0.0029  & 0.0024 & 0.0027  & 0.0035 \\
\hline 
\end{tabular}   
\label{tab:syst_uncert}
\end{center}}
\end{table}
The systematic uncertainties are summarized in Table \ref{tab:syst_uncert}, where the total is the sum in quadrature of the individual contributions. The results for the integrated \CP asymmetries are
\begin{align}
\acp(\kpipi)  = +0.011 \pm 0.002 \pm 0.003 \pm 0.003, \nonumber \\
\acp(\kkk)    = -0.037 \pm 0.002 \pm 0.002 \pm 0.003, \nonumber \\
\acp(\pipipi) = +0.080 \pm 0.004 \pm 0.003 \pm 0.003, \nonumber \\
\acp(\kkpi)   = -0.114 \pm 0.007 \pm 0.003 \pm 0.003, \nonumber
\end{align} 
where the first uncertainty is statistical, the second is systematic and the third is due to the limited knowledge of the \CP asymmetry of the \jpsik control channel~\cite{PDG2020}. 

The significance of the \CP asymmetries is computed by dividing the central values by the sum in quadrature of the uncertainties, yielding  8.5 standard deviations ($\sigma$) for \kkk decays, 14.1$\sigma$ for \pipipi decays and 13.6$\sigma$ for \kkpi decays. This is the first observation of \CP asymmetries in these channels. For the \kpipi decays, the significance of the \CP asymmetry is 2.4$\sigma$, consistent with \CP conservation.


\section{\boldmath Localised \CP asymmetries}
\label{sec:regional}

\begin{figure}[tb]
\centering
\begin{overpic}[width=0.49\linewidth]{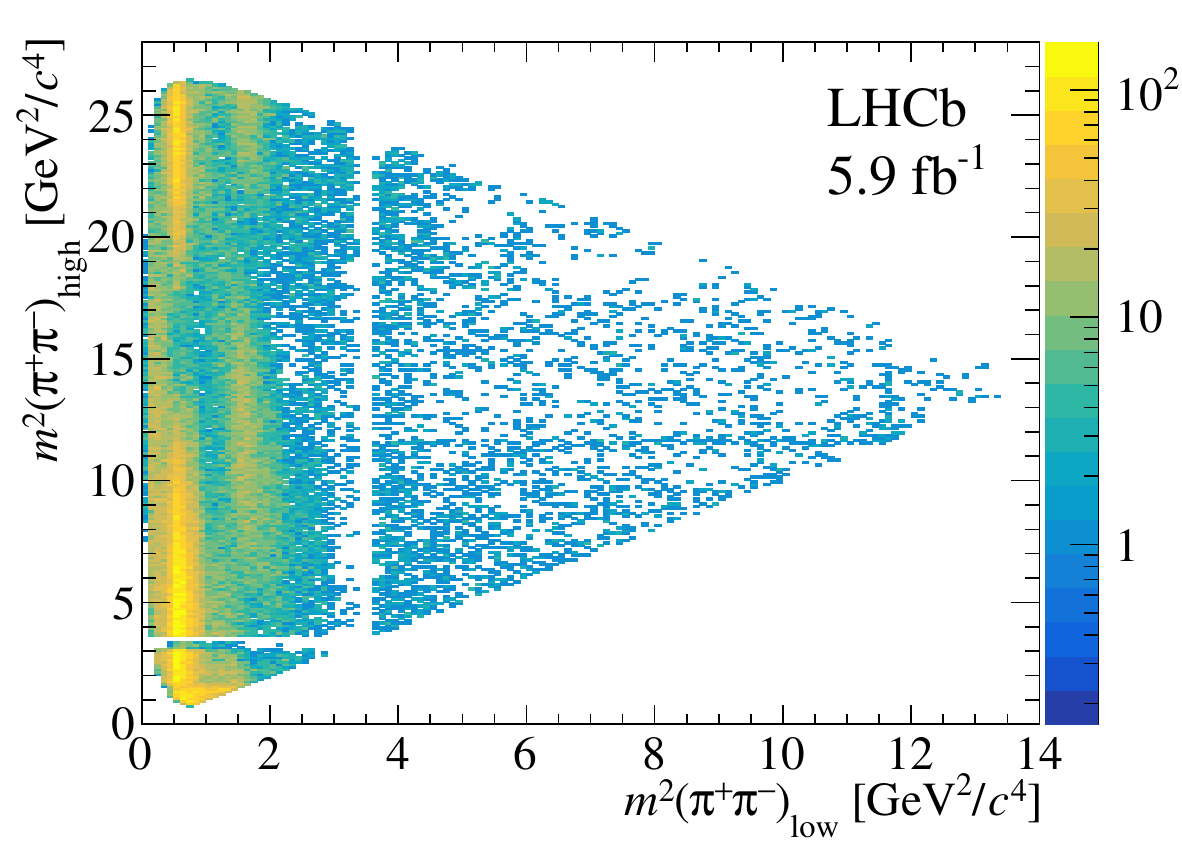}
 \put(50,64){\scriptsize{(a)}}
\end{overpic}
\begin{overpic}[width=0.49\linewidth]{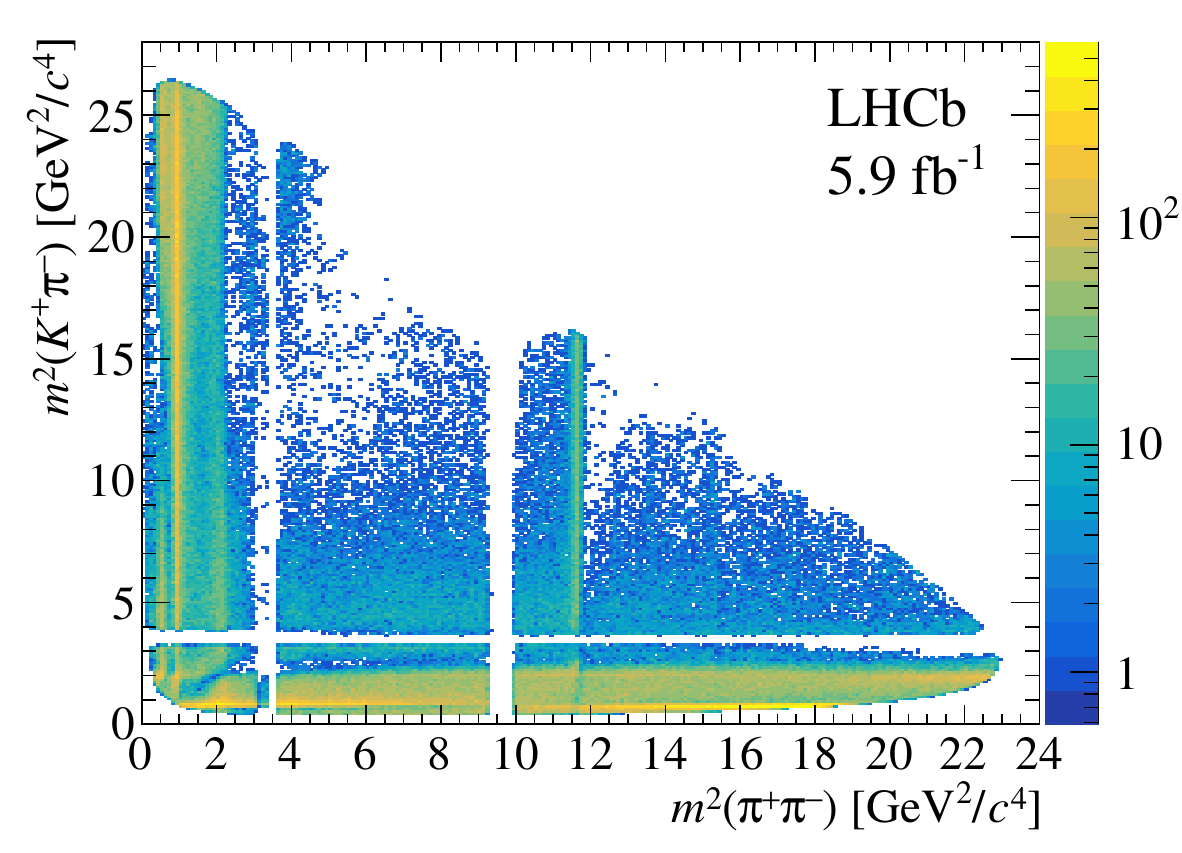}
 \put(50,64){\scriptsize{(b)}}
\end{overpic}

\vspace{0.2cm}

\begin{overpic}[width=0.49\linewidth]{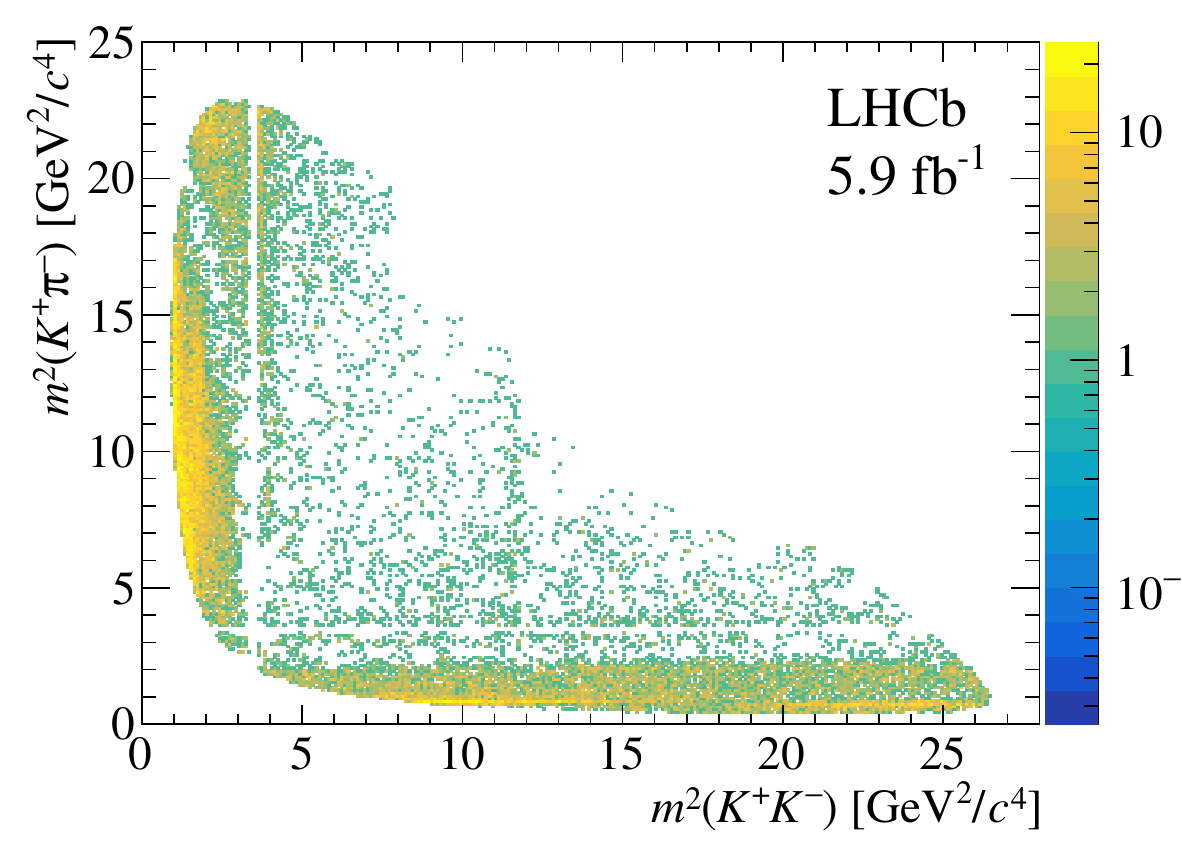}
 \put(50,64){\scriptsize{(c)}}
\end{overpic}
\begin{overpic}[width=0.49\linewidth]{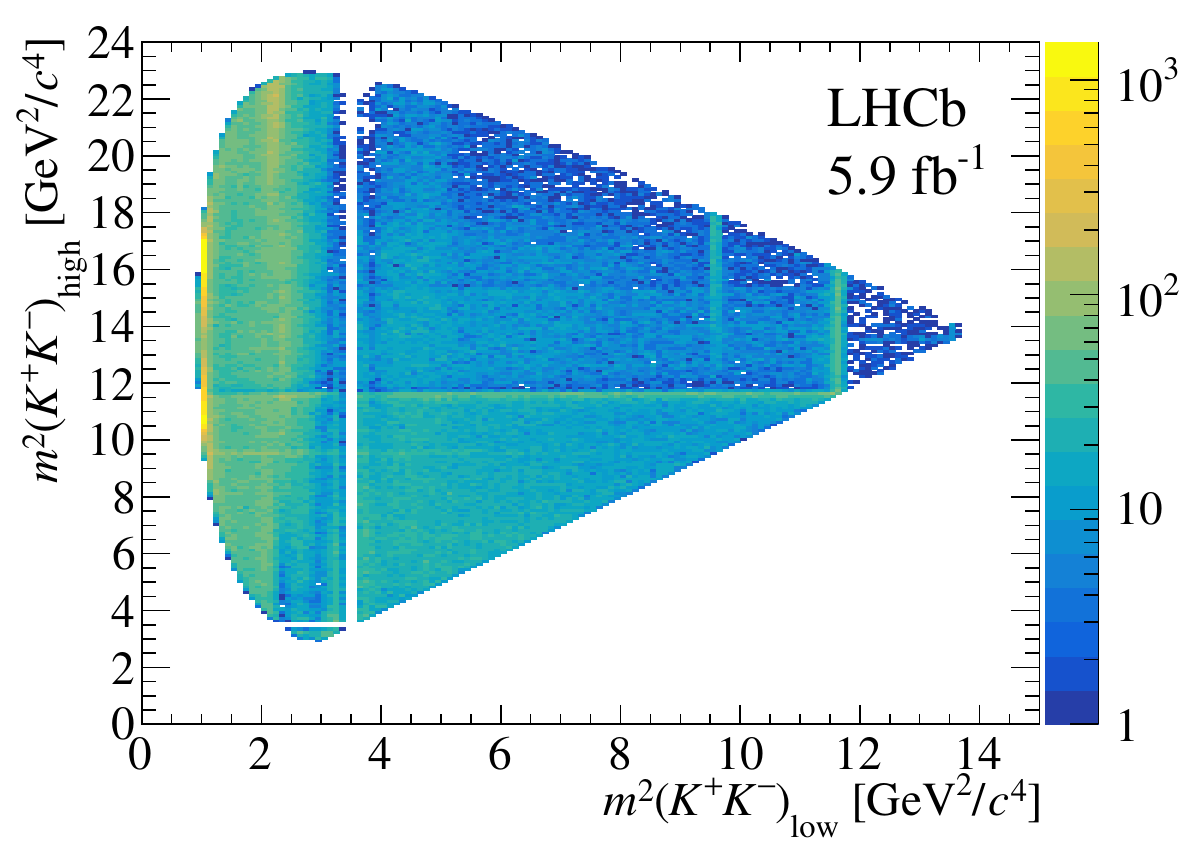}
  \put(50,64){\scriptsize{(d)}}
\end{overpic}
\caption{Dalitz plot distributions for (a)\,\pipipi, (b)\,\kpipi, (c)\,\kkpi and (d)\,\kkk decays. The colour scale indicates the number of events.
}
\label{fig:Dalitz}
\end{figure}

The two-dimensional phase space referred to as the Dalitz plot, formed by the invariant masses squared of two of the three possible particle pairs, reflects directly the dynamics of the decay and allows the study of the different resonant and nonresonant components to inspect \CP asymmetry effects. 
The Dalitz plot distributions of signal yield, obtained with the \sPlot technique \cite{Pivk:2004ty}, for \pipipi, \kpipi, \kkpi and \kkk are shown in Fig.~\ref{fig:Dalitz}. In the symmetric channels, the phase space distribution and its projections are presented with the two axes being the squares of the low-mass $m_{\textrm{low}}$ and high-mass $m_{\textrm{high}}$ combinations of the opposite-sign particle pairs, for visualization purposes.

Most of the candidates are concentrated in the low-mass regions, as expected for charmless decays dominated by resonant contributions. The gap from the vetoed potential $J/\psi$ contributions is visible in the \kpipi channel, as well as the gaps from $D^0$ regions excluded in all modes.

\begin{figure}[tb]
\begin{overpic}[width=0.49\linewidth]{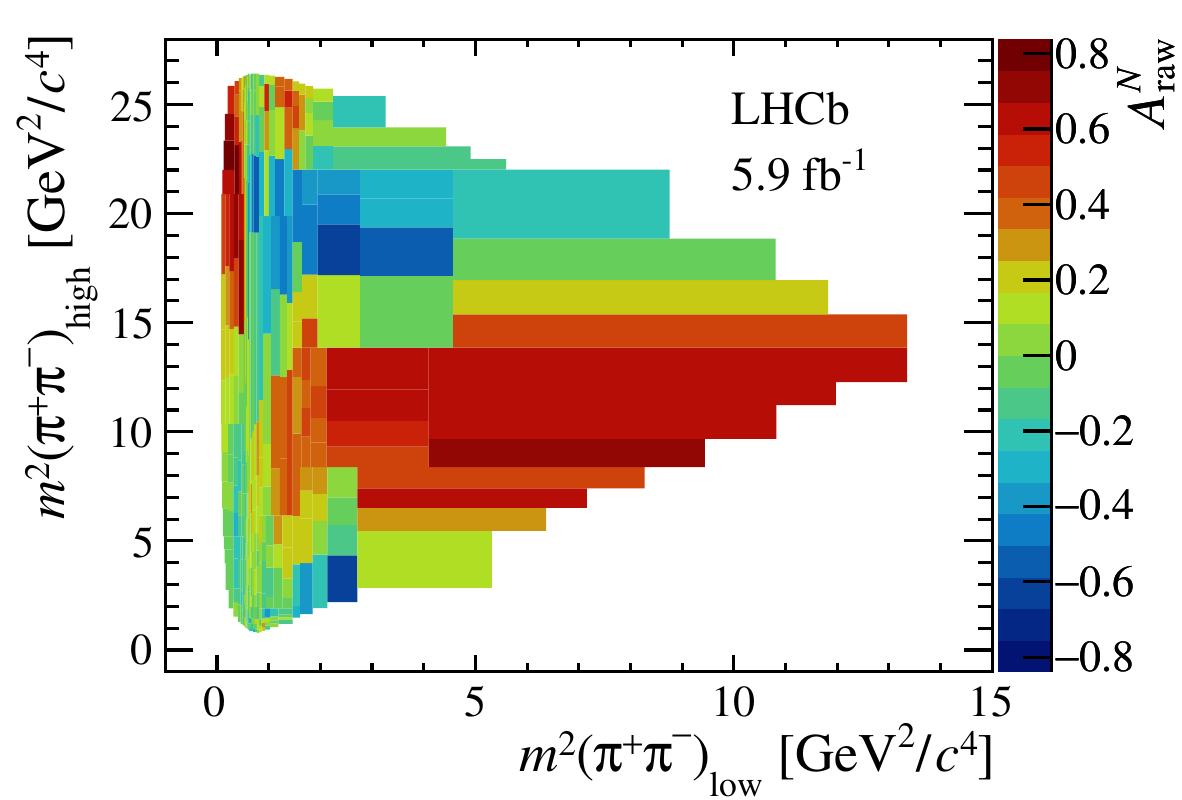}
 \put(74,59){\scriptsize{(a)}}
\end{overpic}
\begin{overpic}[width=0.49\linewidth]{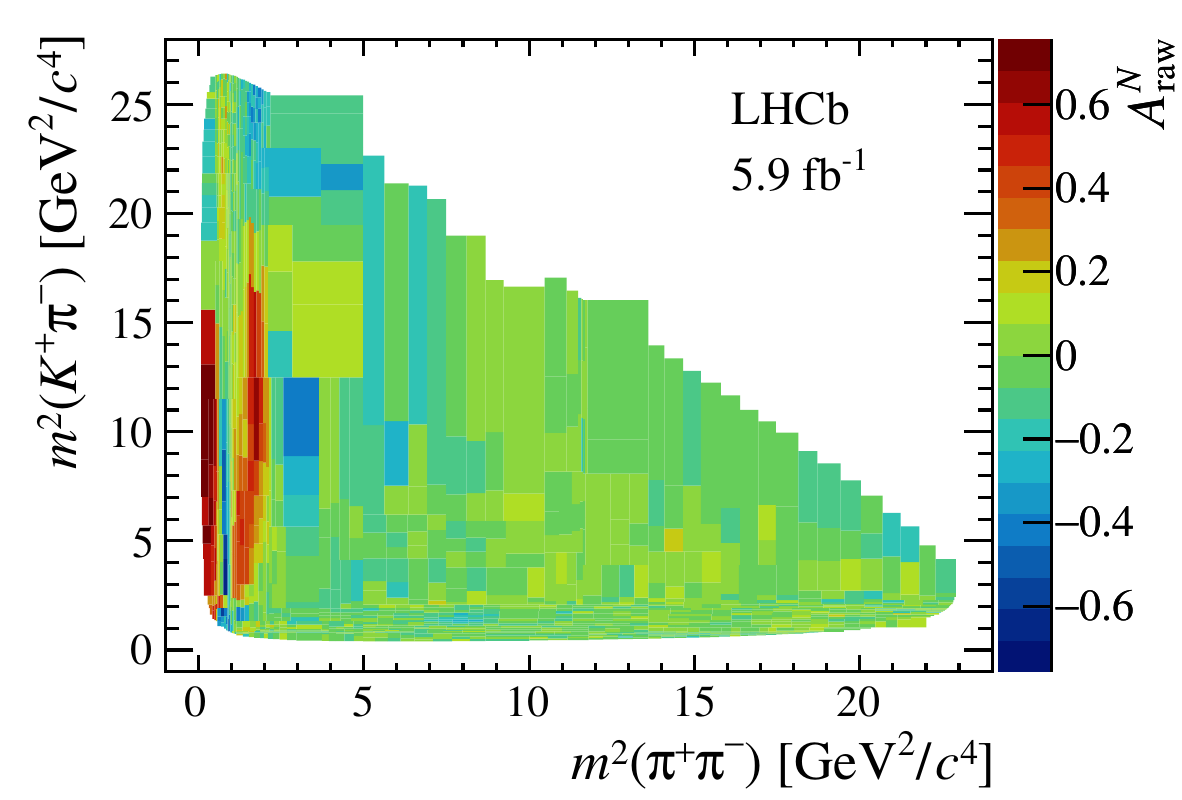}
 \put(74,59){\scriptsize{(b)}}
\end{overpic}

\vspace{0.2cm}

\begin{overpic}[width=0.49\linewidth]{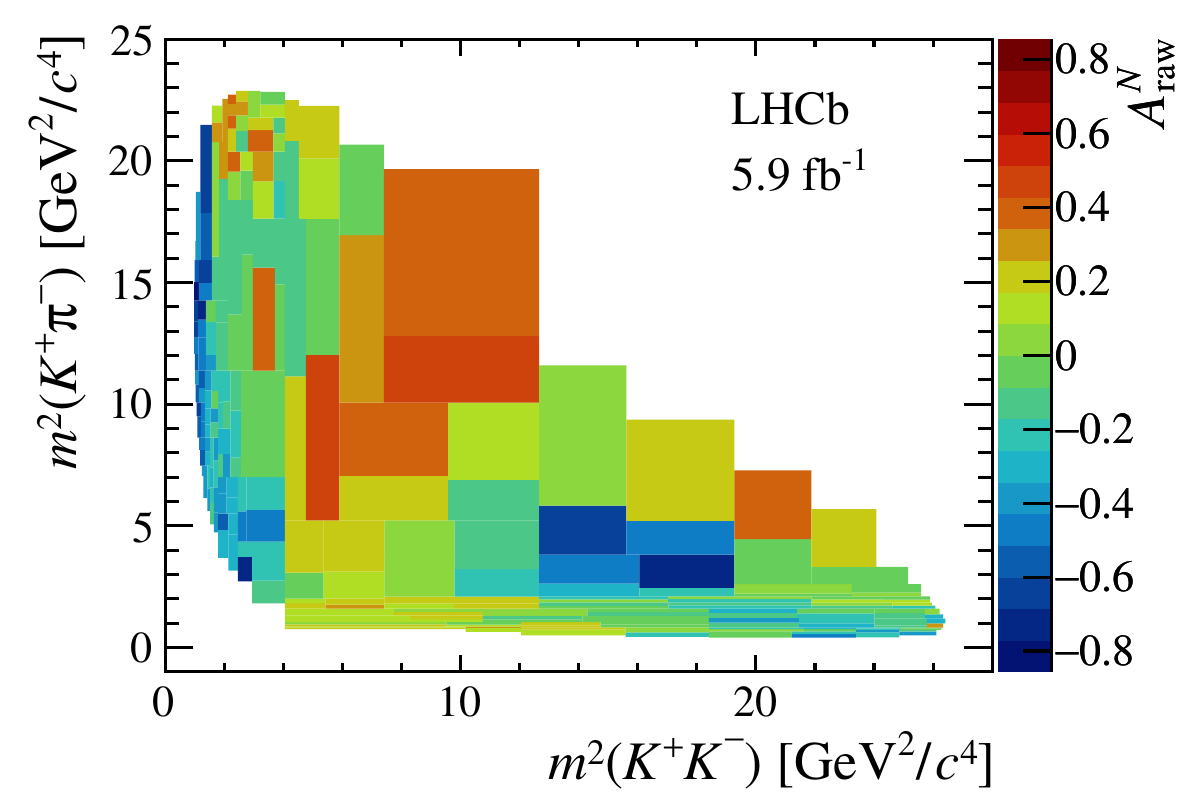}
 \put(74,59){\scriptsize{(c)}}
\end{overpic}
\begin{overpic}[width=0.49\linewidth]{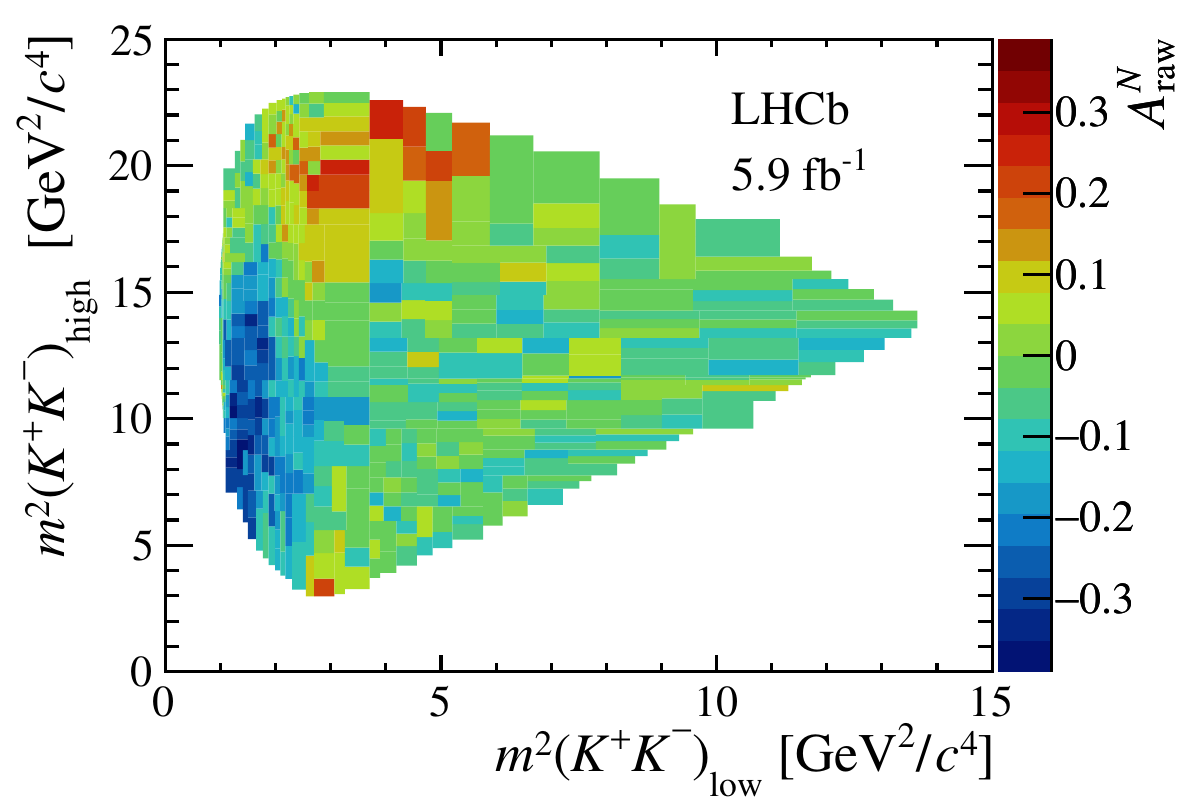}
  \put(74,59){\scriptsize{(d)}}
\end{overpic}

\caption{Asymmetry distribution in bins of the Dalitz plot for (a) \pipipi, using 400 bins with approximately 229 events/bin; (b) \kpipi, using 1728 bins with approximately 276 events/bin; (c) \kkpi, using 256 bins with approximately 127 events/bin; and (d) \kkk, using 729 bins with approximately 461 events/bin.}
\label{fig:Mirandizing}
\end{figure}

In order to visualise localized asymmetries, the \acp in bins of the phase space~\cite{Miranda2} is constructed. Adaptive binning is employed, such that the efficiency-corrected signal yield, also obtained with the \sPlot technique, is approximately equal in all bins. There is no specific rule for choosing the best binning except for requiring a minimum bin occupancy. Figure~\ref{fig:Mirandizing} reveals a rich pattern of large and localized asymmetries which result from interference between the contributions, as well as possible $\pi\pi \to KK$ rescattering, as was observed in the amplitude analyses of Refs.~\cite{LHCb-PAPER-2019-017, LHCb-PAPER-2019-018, LHCb-PAPER-2018-051} and elastic scattering experiments~\cite{Pelaez2018,PhysRevD.22.2595}.

Different regions of the Dalitz plots in Fig.~\ref{fig:Mirandizing} are defined to study the localized \CP asymmetries.
The rescattering region~\cite{Pelaez2018} is defined in the Dalitz plot in the two-kaon or two-pion invariant mass range 1.1--2.25\gevgevcccc for \kkk due to the presence of the $\phi(1020)$ resonance and 1.0--2.25\gevgevcccc for the other three channels, as listed in Table~\ref{table:table_regions}. The data are studied in intervals of the other two-body invariant mass. 
The vertical axis projection depends on the decay channel, with two regions defined for \pipipi, \kpipi and \kkk, and one for \kkpi. There are also two other regions covering higher masses for \pipipi and \kkpi decays. All regions for the different decay channels are defined in Table~\ref{table:table_regions}.
The invariant-mass fits used to compute \acp in each region follow the same fit procedure as described in Sec.~\ref{sec:massfits}, where the efficiency ratios are computed separately per region.

\begin{table}
\centering
\caption{Definitions of phase-space regions in units of \gevgevcccc. In the \kkk channel, the $\phi(1020)$ vector meson is excluded by requiring $m^2(K^+K^-)_{\textrm{low}} > 1.1\gevgevcccc$.}
\label{table:table_regions}
\begin{tabular}{c c c c}
\hline
& & \pipipi & \\
Region 1 & $\quad 1 < m^2(\pi^+ \pi^-)_{\textrm{low}} <$ 2.25 & and & $      3.5 < m^2(\pi^+ \pi^-)_{\textrm{high}} < 16$ \\
Region 2 & $\quad 1 < m^2(\pi^+ \pi^-)_{\textrm{low}} <$ 2.25 & and & $\;    16  < m^2(\pi^+ \pi^-)_{\textrm{high}} < 23$ \\
Region 3 & $4       < m^2(\pi^+ \pi^-)_{\textrm{low}} <$ 15   & and & $\quad 4   < m^2(\pi^+ \pi^-)_{\textrm{high}} < 16$ \\
\hline

& & \kpipi & \\
Region 1 & $1 < m^2(\pi^+ \pi^-) < 2.25$ & and & $\; 3.5 < m^2(K^+ \pi^-) < 19.5$ \\
Region 2 & $1 < m^2(\pi^+ \pi^-) < 2.25$ & and & $19.5 < m^2(K^+ \pi^-) < 25.5$ \\
\hline

& & \kkpi & \\
Region 1 & $\quad 1 < m^2(K^+K^-) < 2.25$ & and & $4 < m^2(K^+ \pi^-) < 19$ \\
Region 2 & $4 < m^2(K^+K^-) < 25$         & and & $3 < m^2(K^+ \pi^-) < 16$  \\
\hline

& & \kkk & \\
Region 1 & $1.1 < m^2(K^+ K^-)_{\textrm{low}} < 2.25$  & and & $\; 4 < m^2(K^+ K^-)_{\textrm{high}} < 17$ \\
Region 2 & $1.1 < m^2(K^+ K^-)_{\textrm{low}} < 2.25$  & and & $17 < m^2(K^+ K^-)_{\textrm{high}} < 23$ \\
\hline
\end{tabular}
\end{table}

\begin{figure}[tb]
\centering
\begin{overpic}[width=0.49\linewidth]{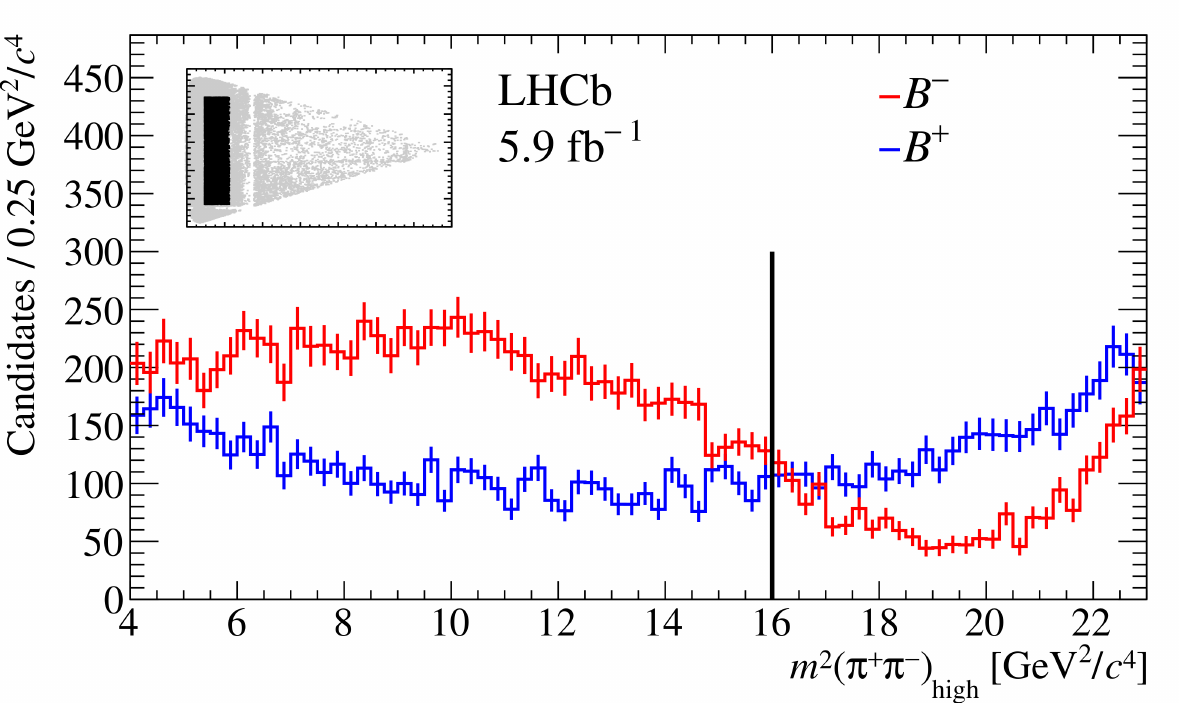}
 \put(90,51){\scriptsize{(a)}}
\end{overpic}

\vspace{0.2cm}

\begin{overpic}[width=0.49\linewidth]{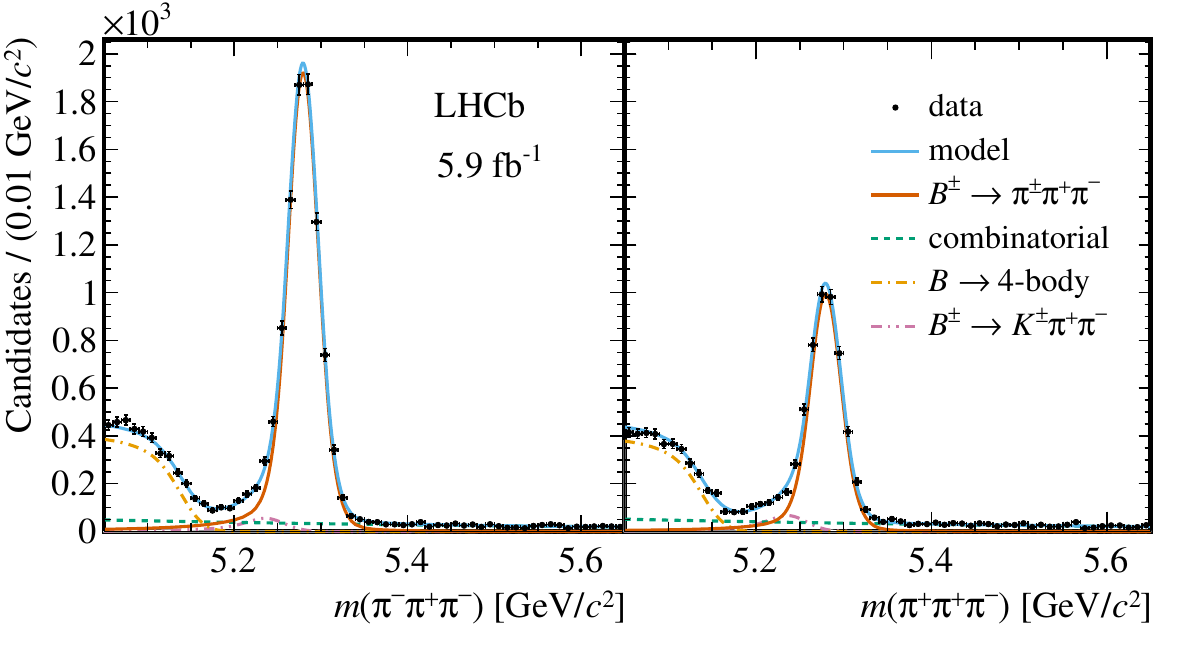}
 \put(90,47){\scriptsize{(b)}}
\end{overpic}
\begin{overpic}[width=0.49\linewidth]{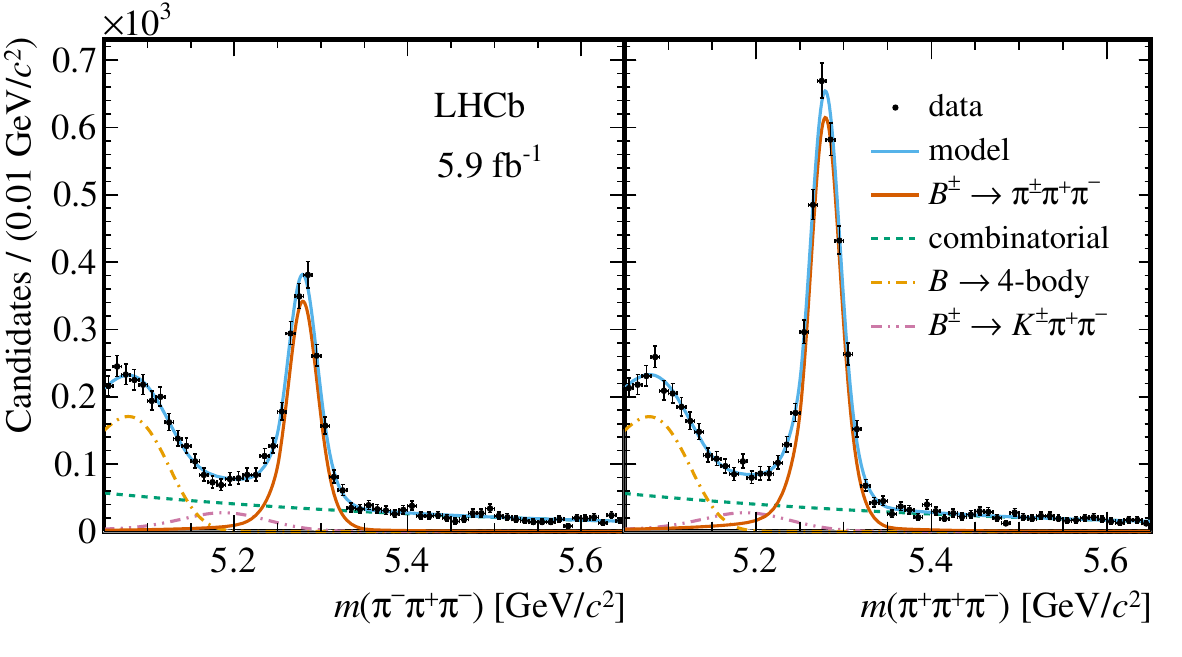}
  \put(90,47){\scriptsize{(c)}}
\end{overpic}
\caption{(a) Projection of $m^2(\pi^+ \pi^-)_{\textrm{high}}$ for the rescattering region (black region inside the Dalitz plot) with the \pipipi mass fits for (b) region 1 and (c) region 2 (\Bm candidates on the left). Regions are separated by a black vertical line in (a), see Table~\ref{table:table_regions}.}
\label{fig:pipipi_region1_results}
\end{figure}

Most of the rescattering regions give \CP asymmetry values in excess of at least five Gaussian standard deviations. For \pipipi decays, the rescattering region when analysed in the $m^2(\pi^+ \pi^-)_{\textrm{high}}$ projection presents a \CP asymmetry that is positive in region~1 and negative in region~2, as shown in Fig.~\ref{fig:pipipi_region1_results}. 
 
\begin{figure}[tb]
\centering
\begin{overpic}[width=0.51\linewidth]{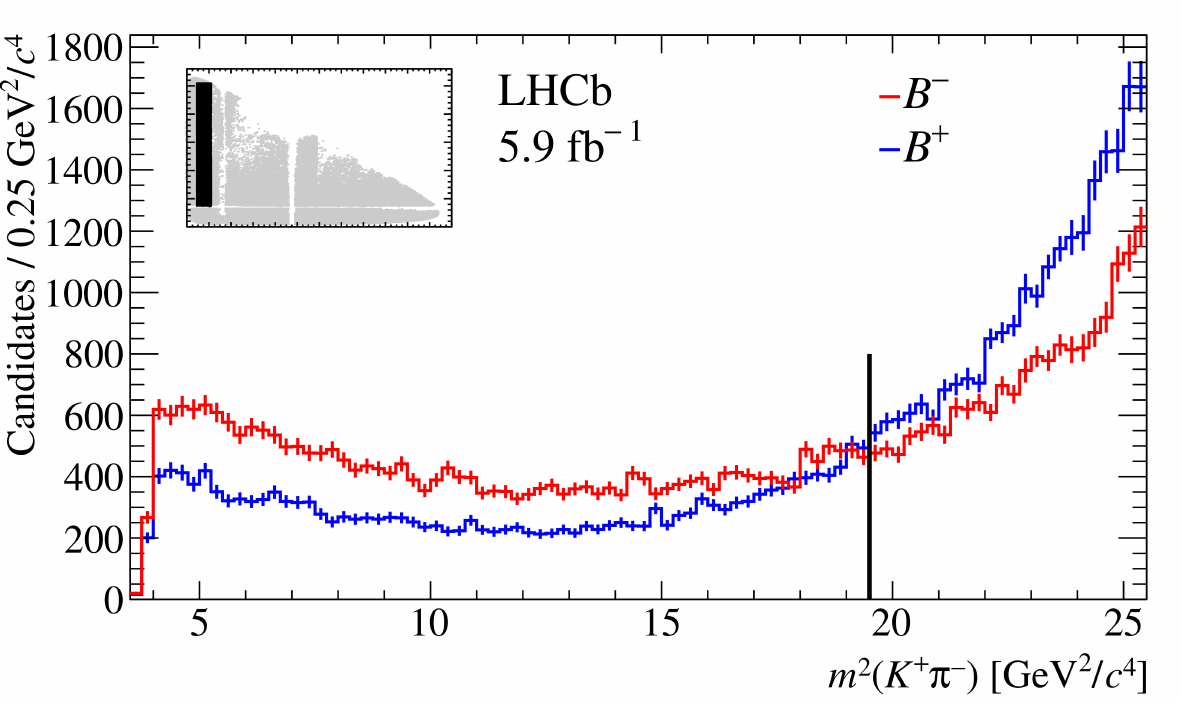}
 \put(88,51){\scriptsize{(a)}}
\end{overpic}

\vspace{0.2cm}

\begin{overpic}[width=0.49\linewidth]{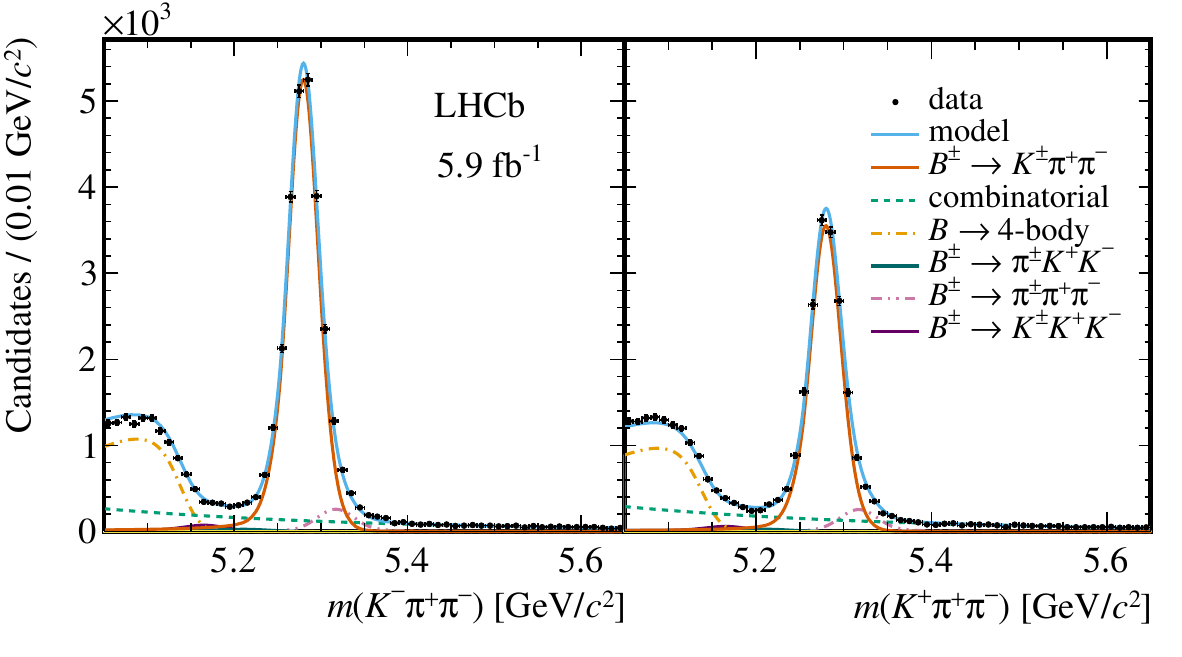}
 \put(90,47){\scriptsize{(b)}}
 \end{overpic}
\begin{overpic}[width=0.49\linewidth]{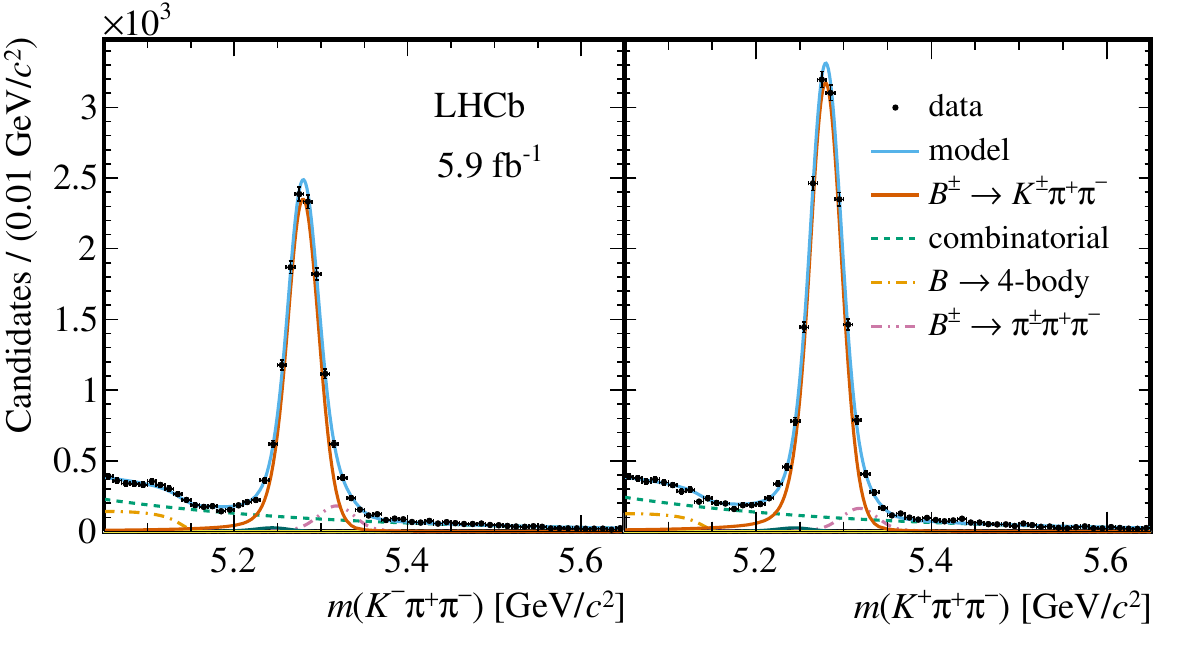}
 \put(90,47){\scriptsize{(c)}}
 \end{overpic}
    \caption{(a) $m^2(K^+ \pi^-)$ projection for the rescattering region with the \kpipi mass fits for (b) region 1 and (c) region 2 (\Bm on the left). Regions are separated by a black vertical line in (a).}
\label{fig:kpipi_region1_results}
\end{figure}

\begin{figure}[tb]
\begin{overpic}[width=0.47\linewidth]{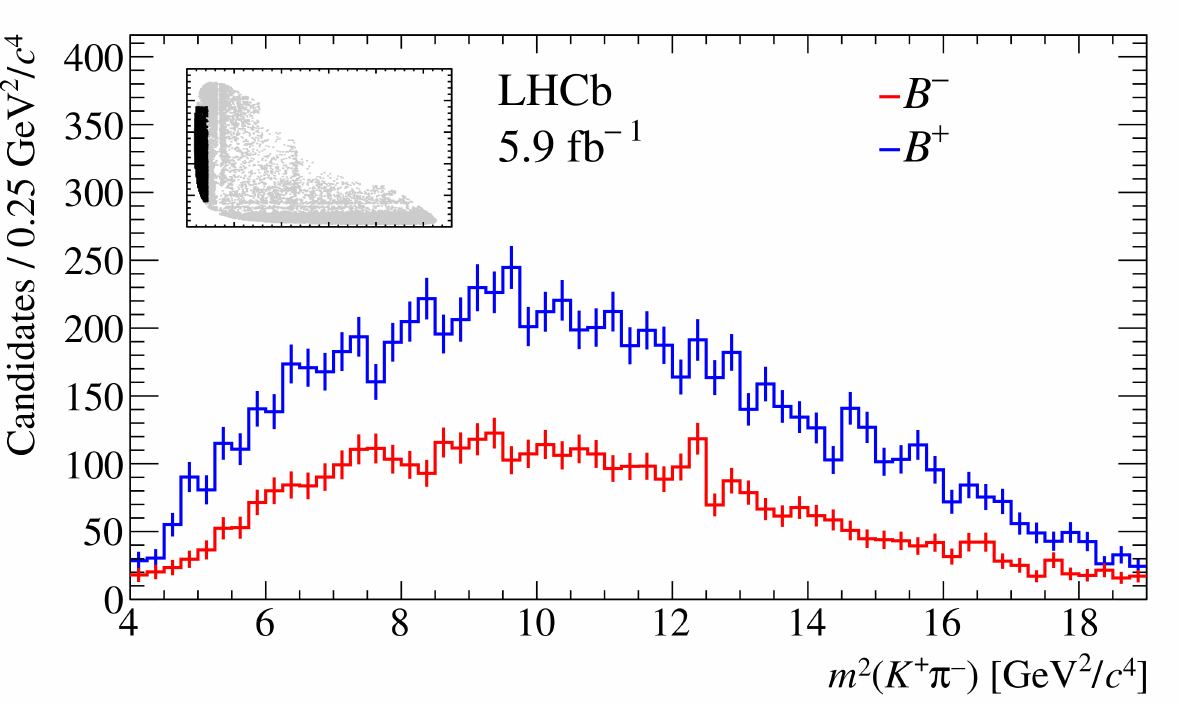}
 \put(88,51){\scriptsize{(a)}}
\end{overpic}
\begin{overpic}[width=0.51\linewidth]{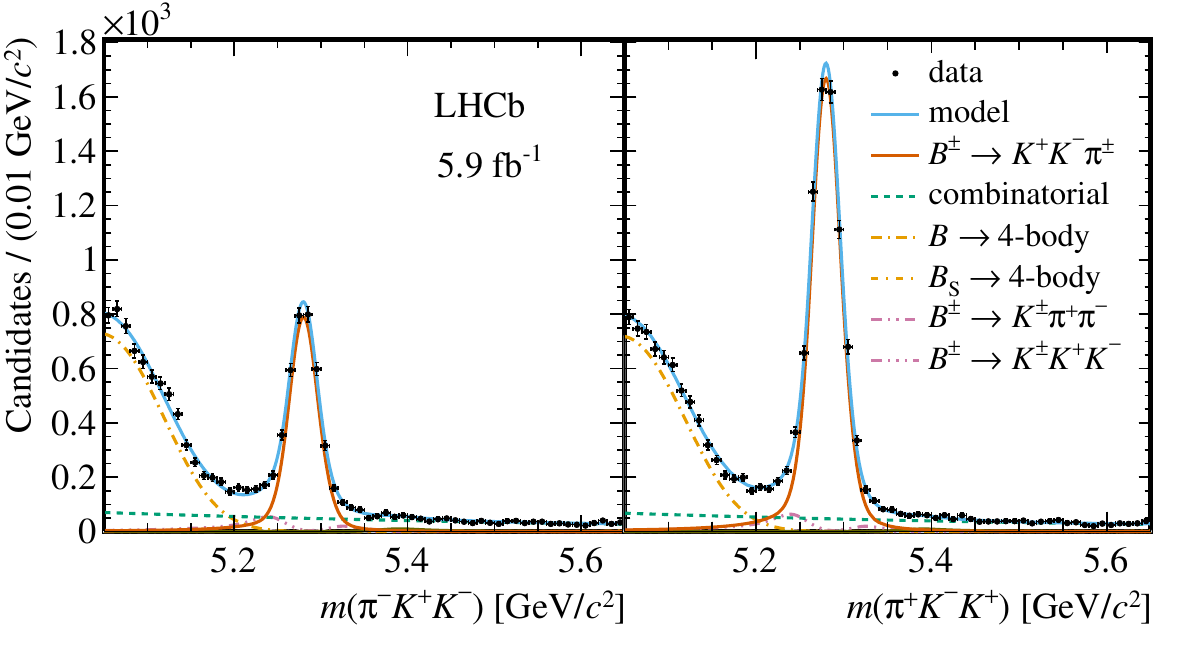}
 \put(90,47){\scriptsize{(b)}}
 \end{overpic}
\caption{(a) $m^2(K^+ \pi^-)$ projection for the rescattering region with the \kkpi mass fits for (b) region 1 (\Bm on the left). }
\label{fig:kkpi_region1_results}
\end{figure}

\begin{figure}[tb]
\centering
\begin{overpic}[width=0.51\linewidth]{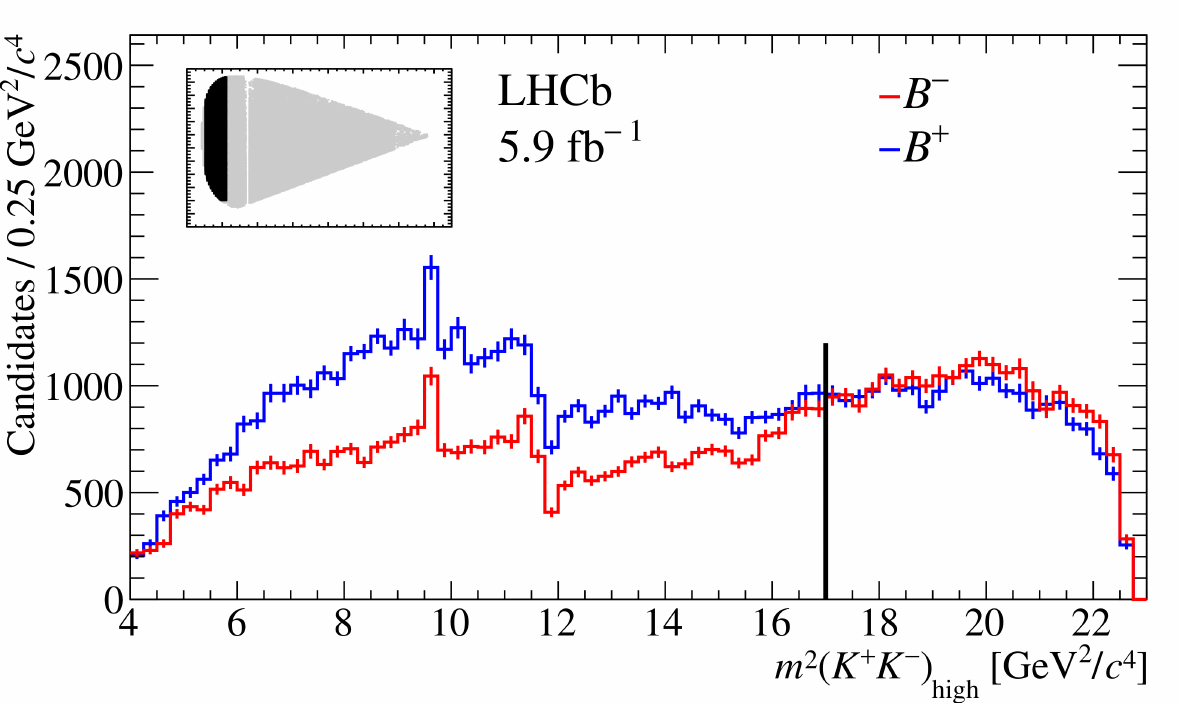}
 \put(88,51){\scriptsize{(a)}}
\end{overpic}

\vspace{0.2cm}

\begin{overpic}[width=0.49\linewidth]{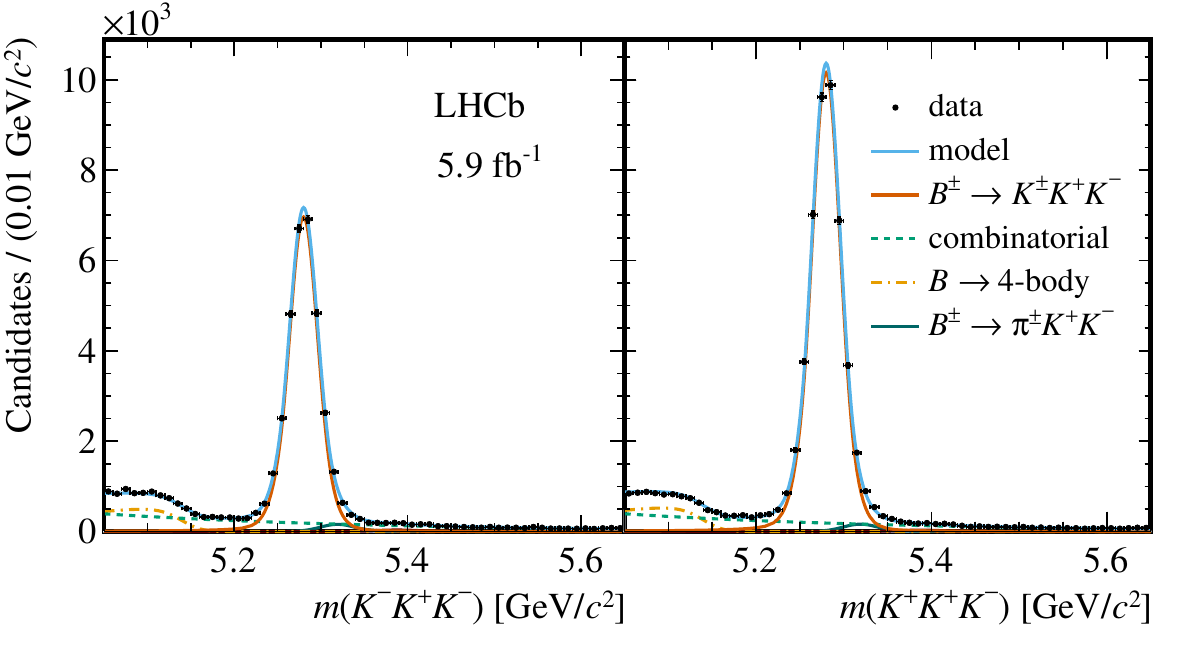}
 \put(90,47){\scriptsize{(b)}}
 \end{overpic}
\begin{overpic}[width=0.49\linewidth]{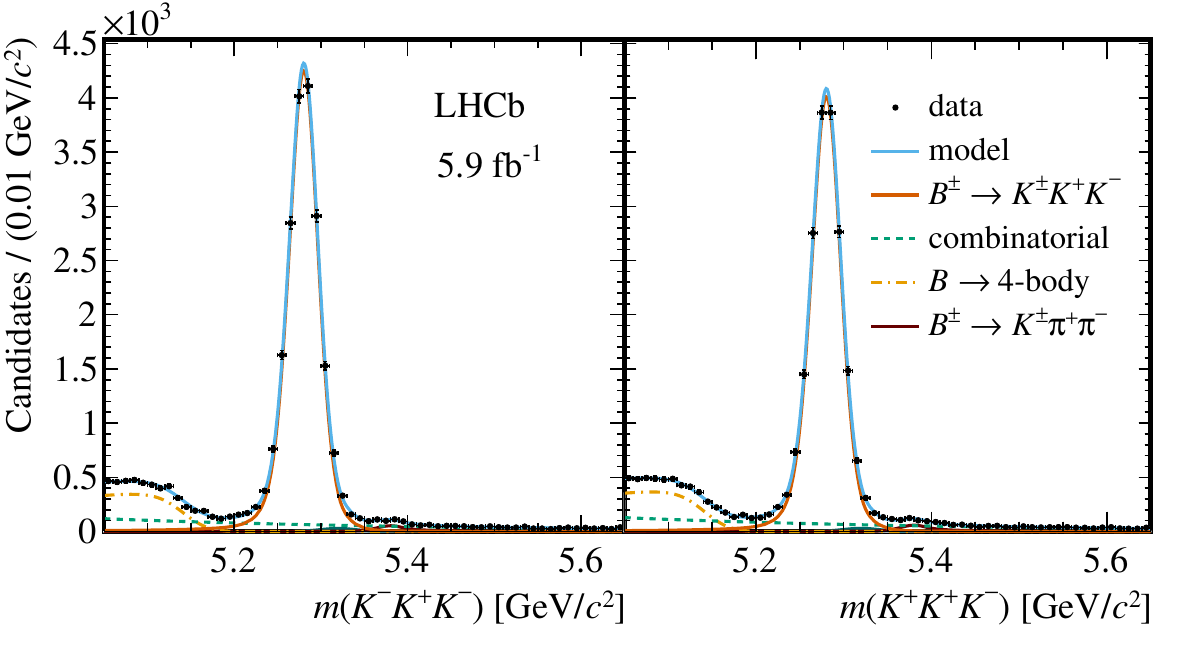}
 \put(90,47){\scriptsize{(c)}}
 \end{overpic}
\caption{(a) $m^2(K^+ K^-)_{\textrm{high}}$ projection for the rescattering region with the \kkk mass fits for (b) region 1 and (c) region 2 (\Bm on the left). Regions are separated by a black vertical line in (a).}
\label{fig:kkk_region1_results}
\end{figure}

For \kpipi decays, the two regions in the $m^2(K^+ \pi^-)$ axis reveal a flip of the asymmetry sign as shown in Fig.~\ref{fig:kpipi_region1_results}. Figure~\ref{fig:kkpi_region1_results} illustrates the $m^2(K^+ \pi^-)$ axis projection for the rescattering region for \kkpi decays, where a nearly constant \CP asymmetry is observed. The \kkk decays also reveal \CP asymmetries in the $m^2(K^+ K^-)_{\textrm{high}}$ projections in the rescattering region, as shown in Fig.~\ref{fig:kkk_region1_results}. Here, another change in the asymmetry sign is observed, in an opposite direction with respect to \kpipi decays. Region 2 of the \kkk decay is the only rescattering projection that shows small asymmetry.

\begin{figure}[tb]
\begin{overpic}[width=0.47\linewidth]{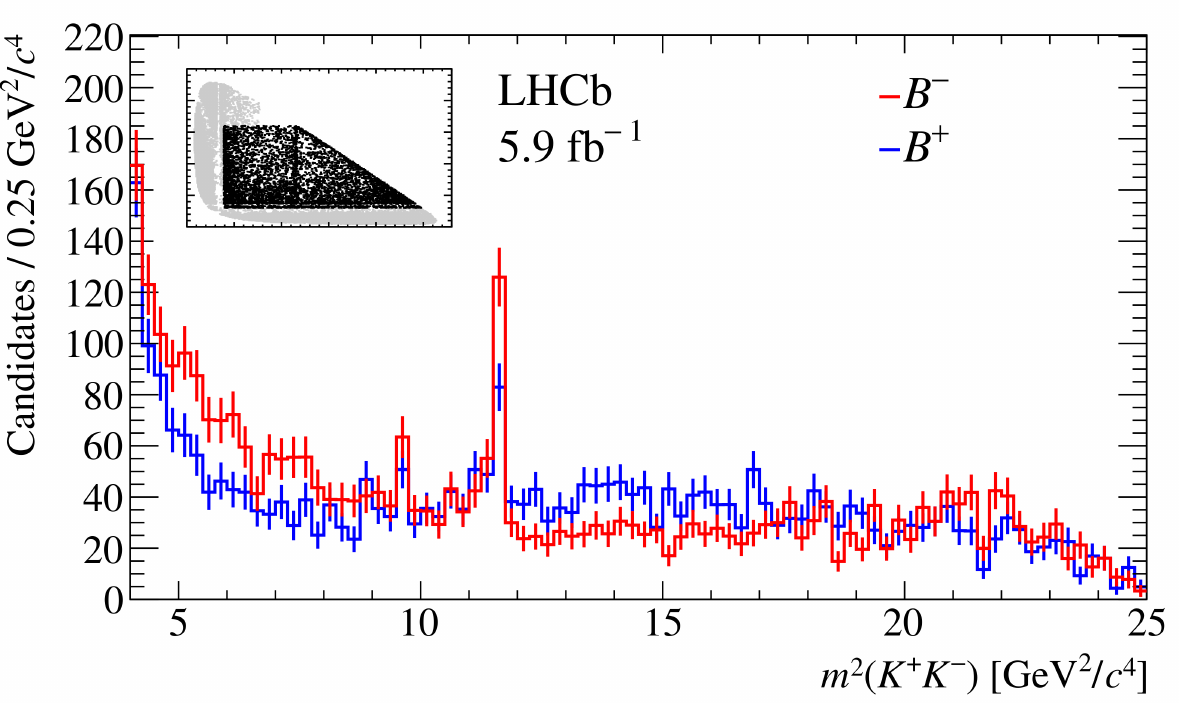}
 \put(88,51){\scriptsize{(a)}}
\end{overpic}
\begin{overpic}[width=0.51\linewidth]{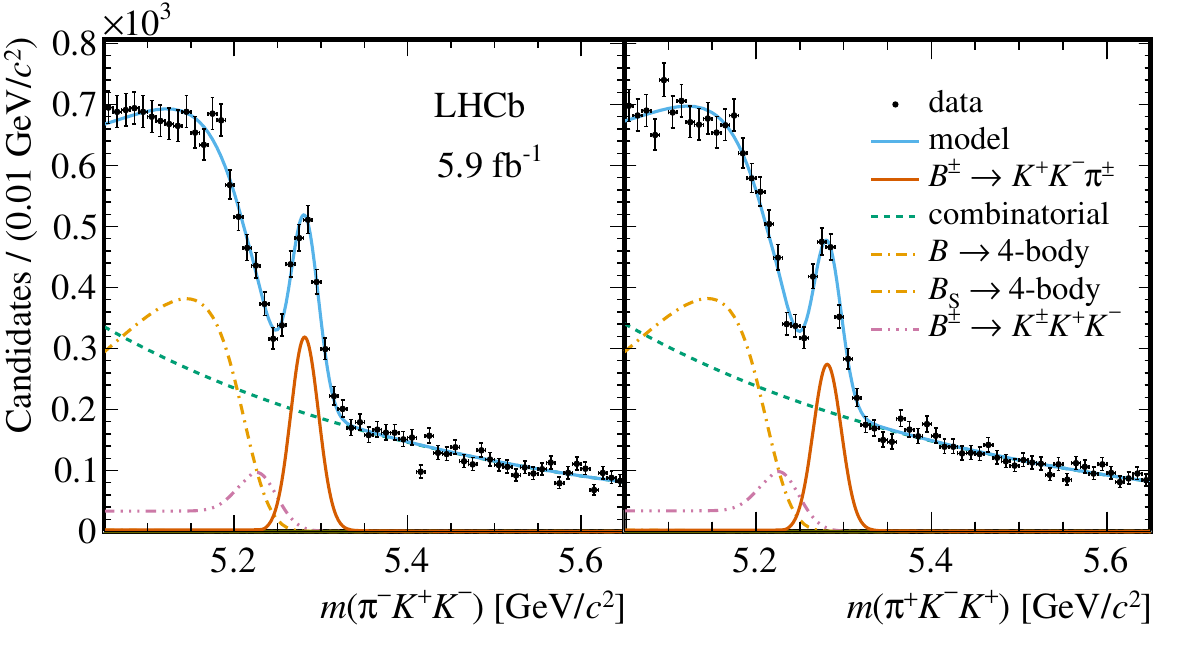}
 \put(90,47){\scriptsize{(b)}}
 \end{overpic}
\caption{(a) $m^2(K^+ K^-)$ projection in the high-mass region (region 2) for \kkpi decays showing the $\chi_{c0}$ contribution. The mass fit for this region (\Bm on the left) is shown in (b).}
\label{fig:kkpi_region2_results}
\end{figure}

\begin{figure}[tb]
\begin{overpic}[width=0.47\linewidth]{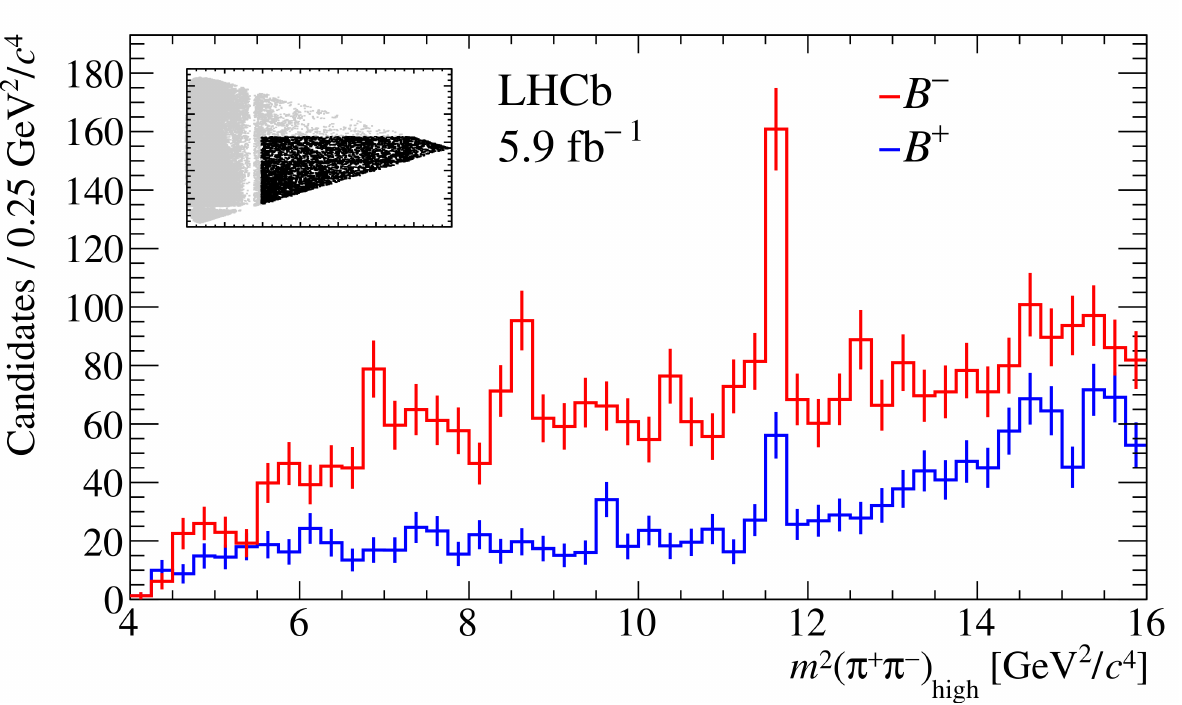}
 \put(88,51){\scriptsize{(a)}}
\end{overpic}
\begin{overpic}[width=0.51\linewidth]{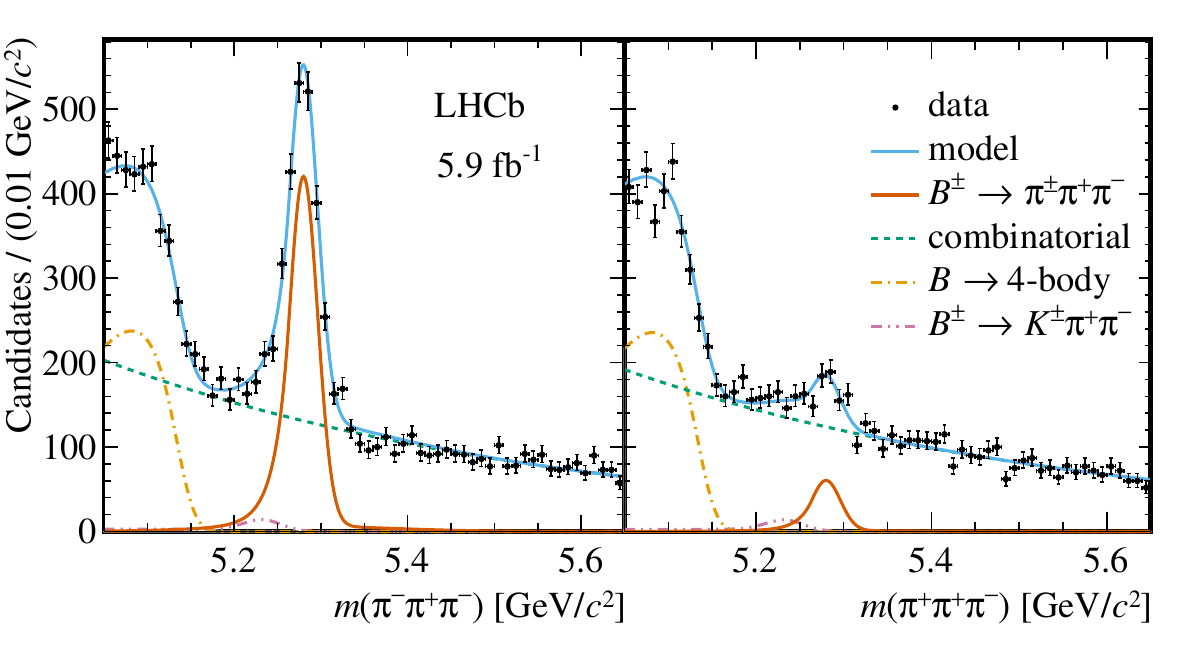}
 \put(90,47){\scriptsize{(b)}}
 \end{overpic}
 \caption{(a) $m^2(\pi^+ \pi^-)_{\textrm{high}}$ projection in the high-mass region (region 3) for \pipipi decays. The mass fit for this region (\Bm on the left) is shown in (b).} 
\label{fig:pipipi_region4_results}
\end{figure}

The high-mass regions of \pipipi and \kkpi as defined in Table~\ref{table:table_regions} are also analysed. For the latter mode, presented in Fig.~\ref{fig:kkpi_region2_results}, no significant asymmetry is observed in the integrated asymmetry. However a clear $\chi_{c0}(1P)$ contribution around $11.6 \gevgevcccc$ is visible. 
For \pipipi decays, a large asymmetry is observed, shown in Fig.~\ref{fig:pipipi_region4_results}, as well as a $\chi_{c0}(1P)$ contribution around $11.6 \gevgevcccc$. The $\chi_{c0}(1P)$ contribution comes from the $B$ decay, as no such structure was observed in the invariant mass sidebands. The asymmetry results from the invariant mass fits for all regions are summarized in Table~\ref{table:table_results_regions}. 

\begin{table}[tb]
\begin{center}
{
\footnotesize
\centering
\caption{Signal yield ($N_{\text{sig}}$), raw asymmetry ($A_{\text{raw}}$) and \acp corrected for production and detection asymmetries. The \acp uncertainties are statistical, systematic and associated to the control channel, respectively. }
\label{table:table_results_regions}

\begin{tabular}{c c c c}
\hline 
\pipipi & $N_{\text{sig}}$ & $A_{\text{raw}}$ & \acp \\
Region 1 & $14\,330 \pm 150$ & $+$0.309 $\pm$ 0.009 & $+$0.303 $\pm$ 0.009 $\pm$ 0.004 $\pm$ 0.003 \\
Region 2 &  $\,\;4\,850 \pm 130$ & $-$0.287 $\pm$ 0.017 & $-$0.284 $\pm$ 0.017 $\pm$ 0.007 $\pm$ 0.003 \\
Region 3 &  $2\,270 \pm 60$  & $+$0.747 $\pm$ 0.027 & $+$0.745 $\pm$ 0.027 $\pm$ 0.018 $\pm$ 0.003 \\

\hline

\kpipi & & & \\
Region 1 & $41\,980 \pm 280$  & $+$0.201 $\pm$ 0.005 & $+$0.217 $\pm$ 0.005 $\pm$ 0.005 $\pm$ 0.003 \\
Region 2 & $27\,040 \pm 250$  & $-$0.149 $\pm$ 0.007 & $-$0.145 $\pm$ 0.007 $\pm$ 0.006 $\pm$ 0.003 \\
\hline

\kkpi & & & \\
Region 1 & $11\,430 \pm 170$  & $-$0.363 $\pm$ 0.010 & $-$0.358 $\pm$ 0.010 $\pm$ 0.014 $\pm$ 0.003 \\
Region 2 & $\,\;2\,600  \pm 120$  & $+$0.075 $\pm$ 0.031 & $+$0.097 $\pm$ 0.031 $\pm$ 0.005 $\pm$ 0.003 \\
\hline

\kkk & & & \\
Region 1 & $76\,020 \pm 350$ & $-$0.189 $\pm$ 0.004 & $-$0.178 $\pm$ 0.004 $\pm$ 0.004 $\pm$ 0.003 \\
Region 2 & $37\,440 \pm 320$ & $+$0.030 $\pm$ 0.005 & $+$0.043 $\pm$ 0.005 $\pm$ 0.004 $\pm$ 0.003 \\
\hline

\end{tabular}
}
\end{center}
\end{table} 

\FloatBarrier


\section{Partial width difference}
\label{sec:uspin}
A relation between \CP violation in \kpipi and  \kkpi, as well as that between  \kkk and  \pipipi, based on U-spin symmetry~\cite{Gronau2003,Gronau2013,Gronau2014}, is investigated. This symmetry is an SU(2) subgroup of the flavour SU(3) group, under which the (\dquark,\squark) pairs of quarks form a doublet, similar to (\uquark,\dquark) in isospin symmetry. 

U-spin symmetry predicts the relationship between the differences of the partial decay widths for the decays \kpipi and \kkpi to be 
\begin{equation}
\Delta\Gamma (\kkpi) = - \Delta\Gamma (\kpipi).
\end{equation} 
A similar prediction is made between the symmetric decay channels \pipipi and \kkk,
\begin{equation}
\Delta\Gamma (\pipipi) = -  \Delta\Gamma (\kkk).
\end{equation} 
These relations state that the difference in the number of events associated to \CP violation must be the same, independently of the branching fraction of each decay.

The above relations can be rewritten in terms of the measured integrated asymmetries, branching fractions and the $B$ meson lifetime. For the \kkpi decay the result is

\begin{equation}
\Delta\Gamma (\kkpi) = \frac{\acp(\kkpi)\mathcal{B}(\kkpi)}{\tau(B^\pm)}\,,
\end{equation} 
where $\tau(B^\pm)$ is the $B^\pm$ lifetime and $\mathcal{B}$ is the average branching fraction. Using the measured values of the integrated asymmetry and branching fractions~\cite{LHCb-PAPER-2020-031}, the U-spin relationship between \kpipi and \kkpi is

\begin{equation} 
\frac{\acp(\kkpi)\mathcal{B}(\kkpi)}{\acp(\kpipi)\mathcal{B}(\kpipi)} = -0.92 \pm 0.18\,. \nonumber
\end{equation} 
Similarly, the pair of \pipipi and \kkk decays give the result

\begin{equation}
\frac{\acp(\pipipi)\mathcal{B}(\pipipi)}{\acp(\kkk)\mathcal{B}(\kkk)} = -1.06 \pm 0.08\,. \nonumber
\end{equation}
Both results are consistent with $-1$ (uncertainties are statistical only), as predicted by the U-spin symmetry approach.
Furthermore, Refs.~\cite{Xu2013, Pat2021} also predicted the relations \hbox{$\Delta\Gamma (\kkpi) = \Delta\Gamma (\kkk)$} and \hbox{$\Delta\Gamma (\kpipi) = \Delta\Gamma (\pipipi)$}. In this work the ratios are measured to be (uncertainties are statistical only):

\begin{align}
\frac{\acp(\kkpi)\mathcal{B}(\kkpi)}{\acp(\kkk)\mathcal{B}(\kkk)} = 0.47 \pm 0.04\,,\nonumber \\
\frac{\acp(\kpipi)\mathcal{B}(\kpipi)}{\acp(\pipipi)\mathcal{B}(\pipipi)} = 0.48 \pm 0.09\,.\nonumber
\end{align} 

The results presented here reveal a central value with an opposite sign with respect to the U-spin prediction of $-1$. However, one may not forget that the U-spin symmetry prediction occurs in a scenario with momentum-independent amplitudes, in which no hadronic final-state interaction where two pairs of channels are coupled through the $\pi\pi \leftrightarrow KK$ interaction is considered. When those contributions are included, one may find ratios that differ from $-1$ without necessarily claiming the observation of U-spin symmetry breaking. All these implications are further discussed in Refs.~\cite{Xu2013, Pat2021}.


\section{Summary and conclusions}
\label{sec:conclusion}

In summary,  the inclusive \CP asymmetries of the \kpipi, \kkpi, \kkk and \pipipi charmless three-body decays are measured with data corresponding to 5.9\invfb of integrated luminosity, collected by the LHCb detector between 2015 and 2018. 
Significant inclusive \CP asymmetries are found for the latter three \B decay channels, the last two observed for the first time. 

The asymmetry in localized regions of phase space shows that the \CP asymmetries are not uniformly distributed in the phase space, with positive and negative \acp appearing in the same charged \B decay channel. 
The results of the previous LHCb analysis~\cite{LHCb-PAPER-2014-044} are confirmed with high localized \CP asymmetries in several regions of the Dalitz plot. It is also confirmed that, in the $\pi\pi \to KK$ rescattering region, between 1 and 2.25\gevgevcccc in both $\pi^+\pi^-$ and $K^+K^-$ invariant masses, significant \CP violation is present in the four analysed channels.
For the \pipipi, \kpipi and \kkk decays, an \acp sign change is observed across the phase space. 

Additionally, an indication of the $\chi_{c0}(1P)$ resonance is seen in the high \pip\pim and \Kp\Km invariant mass regions of \pipipi and \kkpi decays for the first time. There is an indication of \CP violation involving this resonance, as predicted several years ago in Refs.~\cite{Gronau1995,Bediaga1998}.
The combination of the present \acp results with the recent LHCb relative branching fraction results for these decays, shows good agreement with the U-spin symmetry proposed in Refs.~\cite{Gronau2003,Gronau2013,Gronau2014}. 


\section*{Acknowledgements}
%
%
\noindent We express our gratitude to our colleagues in the CERN
accelerator departments for the excellent performance of the LHC. We
thank the technical and administrative staff at the LHCb
institutes.
We acknowledge support from CERN and from the national agencies:
CAPES, CNPq, FAPERJ and FINEP (Brazil); 
MOST and NSFC (China); 
CNRS/IN2P3 (France); 
BMBF, DFG and MPG (Germany); 
INFN (Italy); 
NWO (Netherlands); 
MNiSW and NCN (Poland); 
MEN/IFA (Romania); 
MICINN (Spain); 
SNSF and SER (Switzerland); 
NASU (Ukraine); 
STFC (United Kingdom); 
DOE NP and NSF (USA).
We acknowledge the computing resources that are provided by CERN, IN2P3
(France), KIT and DESY (Germany), INFN (Italy), SURF (Netherlands),
PIC (Spain), GridPP (United Kingdom), 
CSCS (Switzerland), IFIN-HH (Romania), CBPF (Brazil),
Polish WLCG  (Poland) and NERSC (USA).
We are indebted to the communities behind the multiple open-source
software packages on which we depend.
Individual groups or members have received support from
ARC and ARDC (Australia);
Minciencias (Colombia);
AvH Foundation (Germany);
EPLANET, Marie Sk\l{}odowska-Curie Actions and ERC (European Union);
A*MIDEX, ANR, IPhU and Labex P2IO, and R\'{e}gion Auvergne-Rh\^{o}ne-Alpes (France);
Key Research Program of Frontier Sciences of CAS, CAS PIFI, CAS CCEPP, 
Fundamental Research Funds for the Central Universities, 
and Sci. \& Tech. Program of Guangzhou (China);
GVA, XuntaGal and GENCAT (Spain);
SRC (Sweden);
the Leverhulme Trust, the Royal Society
 and UKRI (United Kingdom).




\addcontentsline{toc}{section}{References}
\bibliographystyle{LHCb}
\bibliography{main,standard,LHCb-PAPER,LHCb-DP}
 
\newpage
\centerline
{\large\bf LHCb collaboration}
\begin
{flushleft}
\small
R.~Aaij$^{32}$\lhcborcid{0000-0003-0533-1952},
A.S.W.~Abdelmotteleb$^{50}$\lhcborcid{0000-0001-7905-0542},
C.~Abellan~Beteta$^{44}$,
F.~Abudin{\'e}n$^{50}$\lhcborcid{0000-0002-6737-3528},
T.~Ackernley$^{54}$\lhcborcid{0000-0002-5951-3498},
B.~Adeva$^{40}$\lhcborcid{0000-0001-9756-3712},
M.~Adinolfi$^{48}$\lhcborcid{0000-0002-1326-1264},
H.~Afsharnia$^{9}$,
C.~Agapopoulou$^{13}$\lhcborcid{0000-0002-2368-0147},
C.A.~Aidala$^{77}$\lhcborcid{0000-0001-9540-4988}
% ...

{\footnotesize \it

% ...
$^{9}$Universit{\'e} Clermont Auvergne, CNRS/IN2P3, LPC, Clermont-Ferrand, France\\
% ...
$^{13}$LPNHE, Sorbonne Universit{\'e}, Paris Diderot Sorbonne Paris Cit{\'e}, CNRS/IN2P3, Paris, France\\
% ...
$^{32}$Nikhef National Institute for Subatomic Physics, Amsterdam, Netherlands\\
% ...
$^{40}$Instituto Galego de F{\'\i}sica de Altas Enerx{\'\i}as (IGFAE), Universidade de Santiago de Compostela, Santiago de Compostela, Spain\\
% ...
$^{44}$Physik-Institut, Universit{\"a}t Z{\"u}rich, Z{\"u}rich, Switzerland\\
% ...
$^{48}$H.H. Wills Physics Laboratory, University of Bristol, Bristol, United Kingdom\\
% ...
$^{50}$Department of Physics, University of Warwick, Coventry, United Kingdom\\
% ...
$^{54}$Oliver Lodge Laboratory, University of Liverpool, Liverpool, United Kingdom\\
% ...
$^{62}$Syracuse University, Syracuse, NY, United States\\
% ...
$^{77}$University of Michigan, Ann Arbor, MI, United States, associated to $^{62}$\\
\bigskip
% ...
}
\end{flushleft}

\end{document}